\documentclass[10pt,a4paper]{article}
\usepackage{amssymb,amsmath,amscd,amsthm}
\usepackage{txfonts}
\usepackage{bbm}
\usepackage[dvipdfmx]{graphicx}
\usepackage{braket}

\usepackage{dcolumn}
\usepackage{amsfonts}
\usepackage{bbm} 
\newcommand{\QED}{$\blacksquare$}

\newtheorem{theorem}{Theorem}
\newtheorem{lmm}[theorem]{Lemma}
\newtheorem{cor}[theorem]{Corollary}
\newtheorem{pro}[theorem]{Proposition}
\newtheorem{df}[theorem]{Definition}
\newtheorem{rmk}[theorem]{Remark}

\newcommand\calD{{\cal D}}

\newcommand\calF{{\cal F}}

\newcommand\calK{{\cal K}}

\newcommand\calN{{\cal N}}

\newcommand\calW{{\cal W}}

\newcommand\frah{{\frak h}}
\newcommand\frakk{{\frak k}}
\newcommand\fraH{{\frak H}}
\newcommand\fraK{{\frak K}}

\newcommand\bbR{{\mathbb R}}
\newcommand\bbN{{\mathbb N}}
\newcommand\bbZ{{\mathbb Z}}
\newcommand\bbC{{\mathbb C}}
\newcommand\bbT{{\mathbb T}}
\newcommand\bbH{{\mathbb H}}

\newcommand{\Lvector}[1]{{\left(
\begin{array}{c}
#1 
\end{array} 
\right)}}

\newcommand\stlim{\mathop{\mathrm{st\text{-}lim}}}

\newcommand{\abs}[1]{{\left|#1\right|}}
\newcommand{\norm}[1]{{\left\Vert#1\right\Vert}}

\newcommand{\rbk}[1]{{\left(#1\right)}}
\newcommand{\sbk}[1]{{\left[#1\right]}}
\newcommand{\cbk}[1]{{\left\{#1\right\}}}

\newcommand{\innpro}[1]{{\left\langle#1\right\rangle}}

\begin{document}

\newpage\thispagestyle{empty}
\begin{center}
{\huge\bf
A model of Josephson Junctions \\
on Boson Systems \\
- Currents and Entropy Production Rate -}
\\
\bigskip\bigskip
\bigskip\bigskip
{\Large Tomohiro Kanda}
\\
Graduate School of Mathematics, Kyushu University,
\\
744 Motoka, Nishi-ku, Fukuoka 819-0395, JAPAN
\\
 t-kanda@math.kyushu-u.ac.jp
\\
\end{center}
\bigskip\bigskip\bigskip\bigskip
\bigskip\bigskip\bigskip\bigskip
{\bf Abstract:}
Non-equilibrium steady states (NESS), in the sense of D. Ruelle {\rm [Comm. Math. Phys. {\bf 224}, 3--16 (2001)]}, of Boson systems with Bose--Einstein condensation (BEC) are investigated with the aid of the ${\rm C}^*$-algebraic method.
The model consists of a quantum particle and several bosonic reservoirs.
We show that the mean entropy production rate is strictly positive, independent of phase differences provided that the temperatures or the chemical potentials of reservoirs are different.
Moreover, Josephson currents occur without entropy production, if the temperatures and the chemical potentials of reservoirs are identical.

\newpage

\setcounter{page}{1}
\setcounter{theorem}{0}
\setcounter{equation}{0}
\section{Introduction} \label{sec:Introduction}
\begin{figure}[h]
\begin{center}
\includegraphics[width=15cm,height=10cm]{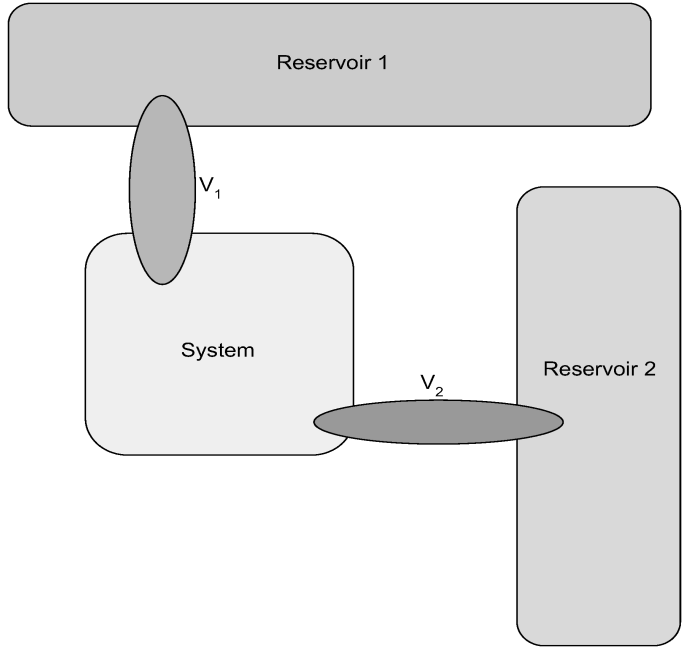}
\caption{Coupled model} \label{fig:Coupled system}
\end{center}
\end{figure}

In the present paper, we study non-equilibrium steady states (NESS) of a model, which consists of a quantum particle and several bosonic reservoirs with Bose--Einstein condensation (BEC) (figure \ref{fig:Coupled system}).
The reservoirs consist of free Bose particles on $\bbR^d$ or on graphs.
We denote the annihilation and the creation operators of the system (resp. the $k$-th reservoir) by $a$ and $a^\dagger$ (resp. $a_{x,k}$ and $a^\dagger_{x,k}$).
These operators satisfy Canonical Commutation Relations (CCR):
\begin{align}
\sbk{a, a^\dagger} = 1, \quad \quad \sbk{ a_{x,k}, a^\dagger_{y,l} } = \delta_{k,l} \delta(x - y), \quad k,l = 1,\ldots, N,
\end{align}
where $N$ is the number of reservoirs.
In the case of $\bbR^d$, the Hamiltonian $H$ of our coupled model is formally given by
\begin{align}
H = H_0 + \lambda \sum_{k=1}^N W_k, \label{eq:General coupled hamiltonian}
\end{align}
where $\lambda > 0$ and
\begin{align}
H_0 = \Omega a^\dagger a + \sum_{k=1}^N \int_{\bbR^d} dp \frac{\abs{p}^2}{2} a^\dagger_{p, k} a_{p,k}, \quad W_k = \int_{\bbR^d} dp \cbk{ \overline{g_k(p)} a^\dagger a_{p,k} + g_k(p) a a^\dagger_{p,k} }. \label{eq:General decoupled hamiltonian and interaction}
\end{align}
When we consider the case of graphs, we replace the integral part of (\ref{eq:General decoupled hamiltonian and interaction}) by the sum over the set of vertices of graphs and $\abs{x}^2/2$ by the adjacency operators of graphs.
Following ideas of D. Ruelle \cite{Ruelle01}, we say that a state is a NESS, if it is a weak $*$-limit point of the net
\begin{align}
\Set{ \frac{1}{T} \int_0^T \omega_0 \circ \alpha_t dt | T > 0},
\end{align}
where $\omega_0$ is the initial state and $\alpha_t$ is the Heisenberg time evolution of our coupled model defined by $\alpha_t(Q) = e^{itH} Q e^{-itH}$ for a quantum observable $Q$.
The initial state is given by a product state of a state of a finite system and equilibrium states with the temperatures, the chemical potentials, and BEC.
The mathematical studies of BEC has a long history (cf. \cite{Verbeure}).
Our line of research is based on a classical paper of J.T. Lewis and J. V. Pul\`{e} \cite{LewisPule} and papers of F. Fidaleo et al. \cite{FidaleoGuidoIsola}, F. Fidaleo \cite{Fidaleo}, and T. Matsui \cite{Matsui2006} .
We obtain explicit formulas of NESS, currents, and the mean entropy production rate.
We prove rigorously that the mean entropy production rate is strictly positive, if the temperatures or the chemical potentials of reservoirs are different and if there exists an open channel, specified in Section \ref{sec:general currents and entropy production rate}.
Moreover, we show that Josephson currents occur without entropy production, if the temperatures and the chemical potentials of reservoirs are identical.

V. Jak\v{s}i\'c, C.-A. Pillet, and their coworkers investigated various aspects of NESS (to take a few examples, \cite{Aschbacher06, AschbacherJaksic, BLP13, BLJP15, Jaksic01, Jaksic02, JaksicOgata}).
However, the case of bosonic reservoirs with BEC was hardly studied before except works of S. Tasaki and T. Matsui \cite{MatsuiTasaki}.

In Section \ref{sec:time evolution}, we have an explicit formula of the coupled time evolution (Theorem \ref{theorem:general Time evolution}).
In Section \ref{sec:NESS}, we define the initial state on the Weyl CCR algebra and obtain an explicit formula of NESS (Theorem \ref{theorem:general Non-equilibrium steady states}).
Section \ref{sec:general currents and entropy production rate} contains our main results: explicit formulas of currents and the (strict) positivity of the mean entropy production rate (Corollary \ref{cor:general Currents}, Proposition \ref{pro:general Entropy production rate}, and Theorem \ref{theorem:positivity of EPR}).
In Section \ref{sec:examples}, we give calculations of currents in the case of $\bbR^d$, $d \geq 3$, and of graphs.
To verify our conditions for the adjacency operators of undirected graphs, we introduce works of M. M\u{a}ntoiu et al. \cite{MantoiuRichardAldecoa}.
They studied the spectrum of the adjacency operators of undirected graphs using Mourre estimate techniques.
After introduction of notations, we consider typical examples of graphs: periodic graphs and comb graphs.

\setcounter{theorem}{0}
\setcounter{equation}{0}
\section{Time Evolutions} \label{sec:time evolution}
In this section, we briefly recall the definition of the Weyl CCR algebras and give an explicit formula of the coupled time evolution.

\subsection{Weyl Operators and Weyl CCR Algebras} \label{sec:Weyl CCR}
Let $\frah$ be a subspace of a Hilbert space $\fraH$.
On the Bose--Fock space $\calF_+(\frah)$, we can define the annihilation operators $a(f)$, and the creation operators $a^\dagger(f)$, $f \in \frah$.
(See e.g. \cite{BratteliRobinsonII}.)
The operators $a(f)$ and $a^\dagger(f)$ are closed and satisfy the equations:
\begin{align}
\sbk{ a(f), a(g)} = 0 = \sbk{a^\dagger(f), a^\dagger(g) }, \quad \sbk{ a(f), a^\dagger(g) } = \innpro{f, g}\mathbbm{1}, \quad f,g \in \frah,
\end{align}
where $\sbk{ A, B } = AB - BA$.
The field operators $\Psi(f)$ are defined by
\begin{align}
\Psi(f) = \frac{1}{\sqrt{2}} \overline{\cbk{ a(f) + a^\dagger(f) }}^{{\rm op. cl.}}, \quad f \in \frah,
\end{align}
where $\overline{A}^{{\rm op. cl.}}$ means the closure of operator $A$.
Then the operators $\Psi(f)$ are (unbounded) self-adjoint and satisfy CCR:
\begin{align}
\sbk{ \Psi(f), \Psi(g) } = {\rm Im} \innpro{f,g} \mathbbm{1} := \sigma(f,g) \mathbbm{1}, \quad f, g \in \frah. \label{eq:CCR}
\end{align}
The Weyl operators $W(f)$ are defined by
\begin{align}
W(f) = \exp(i \Psi(f)), \quad f \in \frah,
\end{align}
and satisfy the following equations:
\begin{align}
W(0) = \mathbbm{1}, \quad W(f)^* = W(-f), \quad W(f)W(g) = e^{ -i \frac{\sigma(f,g)}{2} }W(f + g), \quad f,g \in \frah. \label{eq:Weyl equations}
\end{align}
The Weyl CCR algebra $\calW(\frah)$ is the unital ${\rm C}^*$-algebra generated by unitaries $W(f)$, $f \in \frah$.
Generally, the Weyl CCR algebra $\calW(\frah)$ is the unital universal ${\rm C}^*$-algebra generated by unitaries $W(f)$, $f \in \frah$, which satisfy (\ref{eq:Weyl equations}).
(See e.g. \cite[Theorem 5.2.8.]{BratteliRobinsonII}.)

\subsection{Time Evolutions} 
In this subsection, we give an explicit formula of the coupled time evolutions.
The model is defined on the Boson--Fock space $\calF_+(\calK)$ over the Hilbert space $\calK := \bbC \oplus (\bigoplus_{k=1}^N \fraK_k)$ equipped with the inner product
\begin{align}
\innpro{ \Lvector{c^{(1)} \\ \psi_1^{(1)} \\ \vdots \\ \psi^{(1)}_N}, \Lvector{c^{(2)} \\ \psi_1^{(2)} \\ \vdots \\ \psi^{(2)}_N} } = \overline{c^{(1)}} c^{(2)} + \sum_{k=1}^N \innpro{\psi_k^{(1)}, \psi_k^{(2)}}_k,
\end{align}
where $c^{(1)}, c^{(2)} \in \bbC$, for each $k =1, \ldots, N$, $\fraK_k$ is a Hilbert space with the inner product $\innpro{\cdot, \cdot}_k$, and $\psi_k^{(1)}, \psi_k^{(2)} \in \fraK_k$.
The free Hamiltonian $H_0$ on $\calF_+(\calK)$ is given by $H_0 = d\Gamma(h_0)$, where $d\Gamma$ is the second quantization (see e.g. \cite[Section 5.2]{BratteliRobinsonII}), $h_0$ is the positive self-adjoint operator on $\calK$ defined by
\begin{align}
h_0 \Lvector{c \\ \psi_1 \\ \vdots \\ \psi_N} = \Lvector{ \Omega c \\ h_{0, 1} \psi_1 \\ \vdots \\ h_{0, N} \psi_N }, \label{eq:uncoupled hamiltonian}
\end{align}
$\Omega > 0$, $c \in \bbC$, $h_{0,k}$ is the positive one-particle Hamiltonian on each reservoirs,  $k=1, \ldots,N$, and $\psi_k$ is a vector in the domain of $h_{0,k}$.
The Hamiltonian $H$ of our coupled model is given by $H = d\Gamma(h)$, where $h$ is the self-adjoint operator on $\calK$ defined by
\begin{align}
h  \Lvector{c \\ \psi_1 \\ \vdots \\ \psi_N} =  \Lvector{\Omega c + \lambda \sum_{k=1}^N \innpro{g_k, \psi_k } \\ h_{0,1} \psi_1 + \lambda c g_1 \\ \vdots \\ h_{0, N} \psi_N + \lambda c g_N} =: (h_0 + \lambda V) \Lvector{c \\ \psi_1 \\ \vdots \\ \psi_N}, \label{eq:coupled hamiltonian}
\end{align}
$\lambda > 0$, and $g_k \in \fraK_k$, $k=1, \ldots ,N$.
On the Weyl CCR algebra $\calW(\calK)$, the map $\alpha_t$, $t \in \bbR$, defined by
\begin{align}
\alpha_t(W(f)) = e^{it d\Gamma(h)} W(f) e^{-itd \Gamma(h)} = W(e^{ith} f), \quad f \in \calK,
\end{align}
is a one-parameter group of automorphisms on $\calW(\calK)$.

For simplicity, we denote vectors ${}^t(\psi_1, \ldots, \psi_N)$, ${}^t(g_1, \ldots, g_N)$, and the self-adjoint operator $\bigoplus_{k=1}^N h_{0,k}$ by $\psi$, $g$, and $h_{0,0}$, respectively, where ${}^t(\ldots)$ means transposition.

To obtain an explicit formula of the coupled time evolution, we need some conditions.
\begin{description}
\item[{\rm (Abs)}] For $k=1, \ldots, N$, a pair $(\psi, \xi)$ of vectors $\psi, \xi \in \fraK_k$ satisfy
\begin{align}
\sup_{\nu \in \bbR, \varepsilon > 0} \abs{ \innpro{\psi, (\nu - h_{0,k} \pm i\varepsilon)^{-1} \xi } } < \infty.
\end{align}
For simplicity, if vectors $\psi_k, \xi_k \in \fraK_k$, $k= 1, \ldots, N$, satisfy condition {\rm (Abs)}, then we say that $(\psi, \xi)$ has condition {\rm (Abs)}.

\item[{\rm (A)}] The form factor $g$ defined in (\ref{eq:coupled hamiltonian}) has condition {\rm (Abs)}, i.e., $(g,g)$ has condition {\rm (Abs)}. 
\item[{\rm (B)}] We define the function $\eta(z)$ by 
\begin{align}
\eta(z) := z - \Omega - \lambda^2 \int_{\sigma_0} \frac{1}{ z - \nu } d\innpro{ g, E_0(\nu) g}, \label{eq:definition of eta(z)}
\end{align}
where $E_0$ is the spectral measure of $h_{0,0}$ and $\sigma_0$ is the set of spectrum of $h_{0,0}$. 
Then $1 / \eta_+ \in L^\infty(\bbR)$, where $\eta_+(x) = \lim_{\varepsilon \searrow 0} \eta(x + i\varepsilon)$.
\end{description}

\begin{rmk}
By condition {\rm (A)}, there exists a constant $C_g > 0$ such that
\begin{align}
\sup_{\nu \in \bbR, \varepsilon > 0} \abs{ \innpro{g, (\nu - h_{0,0} \pm i \varepsilon)^{-1} g} } < C_g.
\end{align}
If $\Omega \in \sigma_0$, $\lambda$ is sufficiently small, and there exists a constant $C > 0$ such that
\begin{align}
\frac{ d \innpro{g, E_0(\nu) g} }{d\nu} > C
\end{align}
for a.e. $\nu \in [\Omega - 2 \lambda^2 C_g, \Omega + 2 \lambda^2 C_g]$, then the function $\eta$ satisfies condition {\rm (B)}.
\end{rmk}

We define the sets $\frah_k(g_k)$ and $\frah(g)$ by
\begin{align}
\frah_k(g_k) = \Set{ \psi \in \fraK_k | (\psi, g_k) \text{ has condition {\rm (Abs)}} }, \quad \frah(g) = \Set{ {}^t(\psi_1, \ldots, \psi_N) | \psi_k \in \frah_k(g_k) }. \label{eq:def of abs subset}
\end{align}
For any $c \in \bbC$ and any $\psi \in \frah(g)$, we put $f={}^t (c, \psi)$,
\begin{gather}
F(\nu; f) := c + \lambda \innpro{ g, (\nu - h_{0,0} - i0)^{-1} \psi } \, \rbk{ = c + \lambda \lim_{\varepsilon \searrow 0} \innpro{ g, (\nu - h_{0,0} - i\varepsilon)^{-1} \psi }}, \quad \text{ {\rm a.e.} } \nu \in \bbR, \nonumber\\
\varphi_l(f) := \psi_l + \lambda \frac{F(h_{0,0}; f)}{\eta_-(h_{0,0})} g_l, \quad \varphi(f) := \psi + \lambda \frac{F(h_{0,0}; f)}{\eta_-(h_{0,0})} g. \nonumber
\end{gather}

Let $\fraH$ be a Hilbert space.
For any $\xi, \zeta, \psi \in \fraH$, we set 
\begin{align}
(\xi \otimes \zeta) \psi = \innpro{ \zeta, \psi } \xi,
\end{align}
where $\innpro{\cdot, \cdot}$ is the inner product of $\fraH$.

\begin{pro} \label{pro:resolvent of perturbed hamiltonian}
Let $h_0$ and $h$ be the operators defined in {\rm (\ref{eq:uncoupled hamiltonian})} and {\rm (\ref{eq:coupled hamiltonian})}.
Then we have
\begin{align}
(z - h)^{-1} = (z - h_0)^{-1} + B(z) (z - h_0)^{-1}
\end{align}
for $z \in \bbC$ with ${\rm Im} z \neq 0$, where 
\begin{align}
B(z) =& \lambda^2 \frac{ \innpro{ g, (z - h_{0,0})^{-1} g } }{ \eta(z) } \Lvector{1 \\ 0} \otimes \Lvector{ 1 \\ 0 } + \lambda \frac{z - \Omega}{\eta(z)} \Lvector{ 0 \\ (z - h_{0,0})^{-1} g } \otimes \Lvector{ 1 \\ 0 } \nonumber\\
& + \frac{ \lambda }{ \eta(z) } \Lvector{ 1 \\ 0 } \otimes \Lvector{ 0 \\ g } + \frac{ \lambda^2  }{ \eta(z) } \Lvector{ 0 \\ (z - h_{0,0})^{-1} g } \otimes \Lvector{ 0 \\  g } \label{eq:form of B(z)}
\end{align}
and the function $\eta(z)$ is defined in {\rm (\ref{eq:definition of eta(z)})}.
\end{pro}

\noindent
{\bf Proof.}
By resolvent formula, we have
\begin{align}
(z - h)^{-1} = (z - h_{0})^{-1} + B(z) (z - h_{0})^{-1}. \label{eq:perturbed resolvent}
\end{align}
Since $V$ is a finite rank operator, $B(z)$ has the form of
\begin{align}
B(z) = \xi_1(z) \otimes \Lvector{ 1 \\ 0 } + \xi_2(z) \otimes \Lvector{ 0 \\ g} \label{eq:undecided form of B}
\end{align}
for some $\xi_1(z), \xi_2(z) \in \fraH$.
By multiplying the equation (\ref{eq:perturbed resolvent}) by $z - h$ from the right, we have
\begin{align}
B(z) = \lambda (z - h_{0})^{-1} V + \lambda B(z) (z - h_{0})^{-1} V.
\end{align}
By (\ref{eq:undecided form of B}), we obtain the equation
\begin{align}
\xi_1(z) \otimes \Lvector{ 1 \\ 0 }  + \xi_2(z) \otimes \Lvector{ 0 \\ g} =& \lambda \Lvector{ (z - \Omega)^{-1} \\ 0 } \otimes \Lvector{ 0 \\  g } + \lambda \Lvector{ 0 \\ (z - h_{0,0})^{-1}g} \otimes \Lvector{ 1 \\ 0 } \nonumber\\
& + \lambda \innpro{g, (z - h_{0,0})^{-1} g} \xi_2(z) \otimes \Lvector{1 \\ 0} + \frac{\lambda}{z - \Omega} \xi_1(z) \otimes \Lvector{0 \\ g} \label{eq:equation for xi}
\end{align}
and the forms of $\xi_1(z)$ and $\xi_2(z)$:
\begin{align}
\xi_1(z) =& \lambda \frac{ z - \Omega }{ \eta(z) } \Lvector{ 0 \\ (z - h_{0,0})^{-1} g } + \lambda^2\frac{ \innpro{ g, (z - h_{0,0})^{-1} g } }{ \eta(z) } \Lvector{ 1 \\ 0 }, \\
\xi_2(z) =& \frac{ \lambda }{ \eta(z) } \Lvector{ 1 \\ 0 } + \frac{ \lambda^2  }{ \eta(z) } \Lvector{ 0 \\ (z - h_{0,0})^{-1} g }.
\end{align}
Thus, we get the equation (\ref{eq:form of B(z)}).
\QED

\begin{theorem} \label{theorem:general Time evolution}
Under conditions {\rm (A)} and {\rm (B)}, for any $c, d \in \bbC$ and any $\psi, \xi \in \frah(g)$, which $(\psi, \xi)$ has condition {\rm (Abs)}, $e^{ith}$ has the following form:
\begin{align}
\innpro{ \Lvector{d \\ \xi}, e^{ith}\Lvector{c \\ \psi} } = d c(t) + \innpro{\xi, \psi (t)},
\end{align}
where
\begin{align}
c(t) =& \lambda \innpro{ g, \frac{e^{it h_{0,0}}}{\eta_+(h_{0,0})} \varphi(f) }, \label{eq:time evolution of c} \\
\innpro{\xi, \psi(t)} =& \innpro{\xi, e^{ith_{0,0}} \varphi(f)}  - \lambda^2 \int_{\sigma_0} \frac{e^{it \nu}}{\eta_+(\nu)} \innpro{ \xi,  ( h_{0,0} -\nu - i 0)^{-1} g } d \innpro{g, E_0(\nu) \varphi(f)}.  \label{eq:time evolution of psi}
\end{align}
\end{theorem}

To prove the above theorem, we will use the following lemma.

\begin{lmm} \label{lmm:forms of terms of line_integrals}
Assume that conditions {\rm (A)} and {\rm (B)}. 
For any $R > \norm{ h_{0,0} }$ and any $\zeta, \xi \in \frah(g)$, which $(\zeta, \xi)$ has condition {\rm (Abs)}, we have the following equations:
\begin{gather}
\frac{1}{2 \pi i} \lim_{\varepsilon \searrow 0}  \int_{-R}^R \cbk{\frac{e^{it(x - i\varepsilon)}}{\eta(x - i \varepsilon)} - \frac{e^{it(x + i\varepsilon)}}{\eta(x + i \varepsilon)}} dx = \lambda^2 \innpro{g, \frac{  e^{it h_{0,0}}}{ \abs{ \eta_-(h_{0,0}) }^2 }  g}, \label{eq:First part of time evolution} \\
\frac{1}{2 \pi i} \lim_{\varepsilon \searrow 0} \int_{-R}^R \cbk{ \frac{ e^{it(x -i\varepsilon)} \innpro{ \zeta, (x - h_{0,0} - i\varepsilon)^{-1} \xi } }{ \eta(x - i\varepsilon) } - \frac{ e^{it(x + i\varepsilon)} \innpro{ \zeta, (x - h_{0,0} + i\varepsilon)^{-1} \xi } }{ \eta(x + i\varepsilon) } } dx \nonumber\\
= \innpro{\zeta, \frac{ e^{it h_{0,0}} }{ \eta_+(h_{0,0}) } \xi} + \lambda^2 \int_{\sigma_0} \frac{ e^{it\nu} \innpro{ \zeta, (\nu - h_{0,0} - i0)^{-1} \xi } }{ \abs{\eta_-(\nu)}^2 } d\innpro{ g, E_0(\nu) g } \label{eq:Second part of time evolution1}\\
= \innpro{\zeta, \frac{ e^{it h_{0,0}} }{ \eta_-(h_{0,0}) } \xi} - \lambda^2 \int_{\sigma_0} \frac{ e^{it\nu} \innpro{ \zeta, (h_{0,0} - \nu - i0)^{-1} \xi } }{ \abs{\eta_-(\nu)}^2 } d\innpro{ g, E_0(\nu) g }, \label{eq:Second part of time evolution2}\\
 \frac{1}{2 \pi i} \lim_{\varepsilon \searrow 0} \int_{-R}^R \left\{ \frac{ e^{it (x - i\varepsilon)}}{ \eta(x - i\varepsilon)  } \innpro{ g, (x - h_{0,0} - i\varepsilon)^{-1} \zeta } \innpro{ \xi, (x - h_{0,0} - i \varepsilon)^{-1} g } \right. \nonumber\\
\left. - \frac{ e^{it (x + i\varepsilon)}}{ \eta(x + i\varepsilon)}\innpro{g, (x - h_{0,0} + i\varepsilon)^{-1} \zeta} d\innpro{ \xi,  (x - h_{0,0} + i \varepsilon)^{-1} g } \right\} dx \nonumber
\\
= \int_{\sigma_0} \frac{e^{it \nu}}{\eta_-(\nu)} \innpro{ g, (\nu - h_{0,0} - i 0)^{-1} \zeta } d \innpro{ \xi, E_0(\nu) g } + \int_{\sigma_0} \frac{e^{it \nu}}{\eta_+(\nu)} \innpro{ \xi,  (\nu - h_{0,0} + i 0)^{-1} g } d \innpro{g, E_0(\nu) \zeta} \nonumber\\
 + \int_{\sigma_0} \frac{ \lambda^2 e^{it \nu} }{\abs{\eta_-(\nu)}^2}  \innpro{ g, (\nu - h_{0,0} - i 0)^{-1} \zeta } \innpro{ \xi, (\nu - h_{0,0} + i 0)^{-1} g } d\innpro{ g, E_0(\nu) g }. \label{eq:Third part of time evolution}
\end{gather}
\end{lmm}

\noindent
{\bf Proof.}
Since the equations (\ref{eq:First part of time evolution}), (\ref{eq:Second part of time evolution1}), (\ref{eq:Second part of time evolution2}), and (\ref{eq:Third part of time evolution}) can be shown by similar calculations, we only prove (\ref{eq:Third part of time evolution}).
For the left hand side of the equation (\ref{eq:Third part of time evolution}), we obtain
\begin{align}
& \int_{-R}^R \left\{ \frac{ e^{it (x - i\varepsilon)}}{ \eta(x - i\varepsilon)  } \innpro{ g, (x - h_{0,0} - i\varepsilon)^{-1} \zeta } \innpro{ \xi, (x - h_{0,0} - i \varepsilon)^{-1} g } \right. \nonumber\\
& \left. \quad \quad \quad \quad \quad \quad \quad \quad \quad \quad \quad - \frac{ e^{it (x + i\varepsilon)}}{ \eta(x + i\varepsilon)}\innpro{g, (x - h_{0,0} + i\varepsilon)^{-1} \zeta} \innpro{ \xi,  (x - h_{0,0} + i \varepsilon)^{-1} g } \right\} dx \nonumber\\
=& \int_{-R}^R \frac{e^{itx}}{ \abs{\eta(x - i \varepsilon)}^2 } \left\{ e^{t\varepsilon} \eta(x + i\varepsilon) \innpro{ g, (x - h_{0,0} - i\varepsilon)^{-1} \zeta } \innpro{ \xi, (x - h_{0,0} - i \varepsilon)^{-1} g } \right. \nonumber\\
& \left. \quad \quad \quad \quad \quad \quad \quad \quad \quad \quad \quad - e^{-t\varepsilon} \eta(x - i\varepsilon)\innpro{g, (x - h_{0,0} + i\varepsilon)^{-1} \zeta} \innpro{ \xi,  (x - h_{0,0} + i \varepsilon)^{-1} g } \right\} dx. \label{eq:Third part of time evolution 1}
\end{align}
The integrand in (\ref{eq:Third part of time evolution 1}) has the form of
\begin{align}
& e^{t\varepsilon} \eta(x + i\varepsilon) \innpro{ g, (x - h_{0,0} - i\varepsilon)^{-1} \zeta } \innpro{ \xi, (x - h_{0,0} - i \varepsilon)^{-1} g } \nonumber\\
& - e^{-t\varepsilon} \eta(x - i\varepsilon)\innpro{g, (x - h_{0,0} + i\varepsilon)^{-1} \zeta} \innpro{ \xi,  (x - h_{0,0} + i \varepsilon)^{-1} g } \nonumber\\
=& e^{t\varepsilon} \eta(x + i\varepsilon) \innpro{ g, (x - h_{0,0} - i\varepsilon)^{-1} \zeta } \rbk{ \innpro{ \xi, (x - h_{0,0} - i \varepsilon)^{-1} g } - \innpro{ \xi, (x - h_{0,0} + i \varepsilon)^{-1} g } } \nonumber\\
& + e^{-t\varepsilon} \eta(x - i\varepsilon) \innpro{ \xi,  (x - h_{0,0} + i \varepsilon)^{-1} g } \rbk{ \innpro{g, (x - h_{0,0} - i\varepsilon)^{-1} \zeta} - \innpro{g, (x - h_{0,0} + i\varepsilon)^{-1} \zeta} } \nonumber\\
& + \rbk{ e^{t\varepsilon} \eta(x + i\varepsilon) - e^{-t\varepsilon} \eta(x - i\varepsilon) } \innpro{ g, (x - h_{0,0} - i\varepsilon)^{-1} \zeta } \innpro{ \xi, (x - h_{0,0} + i \varepsilon)^{-1} g }. \label{eq:Third part of integrand}
\end{align}
By conditions {\rm (Abs)}, {\rm (A)}, and {\rm (B)} and (\ref{eq:Third part of integrand}), the limit of (\ref{eq:Third part of time evolution 1}) has the following form:
\begin{align}
& \frac{1}{2 \pi i} \lim_{\varepsilon \searrow 0} \int_{-R}^R \left\{ \frac{ e^{it (x - i\varepsilon)}}{ \eta(x - i\varepsilon)  } \innpro{ g, (x - h_{0,0} - i\varepsilon)^{-1} \zeta } \innpro{ \xi, (x - h_{0,0} - i \varepsilon)^{-1} g } \right. \nonumber\\
& \left. \quad \quad \quad \quad \quad \quad \quad \quad \quad \quad \quad - \frac{ e^{it (x + i\varepsilon)}}{ \eta(x + i\varepsilon)}\innpro{g, (x - h_{0,0} + i\varepsilon)^{-1} \zeta} d\innpro{ \xi,  (x - h_{0,0} + i \varepsilon)^{-1} g } \right\} dx \nonumber\\
=& \int_{\sigma_0} \frac{e^{it \nu}}{\eta_-(\nu)} \innpro{ g, (\nu - h_{0,0} - i 0)^{-1} \zeta } d \innpro{ \xi, E_0(\nu) g } + \int_{\sigma_0} \frac{e^{it \nu}}{\eta_+(\nu)} \innpro{ \xi,  (\nu - h_{0,0} + i 0)^{-1} g } d \innpro{g, E_0(\nu) \zeta} \nonumber\\
& + \int_{\sigma_0} \frac{ \lambda^2 e^{it \nu} }{\abs{\eta_-(\nu)}^2}  \innpro{ g, (\nu - h_{0,0} - i 0)^{-1} \zeta } \innpro{ \xi, (\nu - h_{0,0} + i 0)^{-1} g } d\innpro{ g, E_0(\nu) g }. \nonumber
\end{align}
We have thus proved the lemma.
\QED

\noindent
{\bf Proof of Theorem \ref{theorem:general Time evolution}.}
By Cauchy's integral formula, $e^{ith}$ has the form of
\begin{align}
e^{ith} = \frac{1}{2 \pi i} \stlim_{R \nearrow \infty, \varepsilon \searrow 0} \int_{C_{\varepsilon, R}} \frac{e^{itz}}{ z - h } dz,
\end{align}
where $\stlim$ is the strong limit, $R > \norm{h_{0,0}}$, $\varepsilon > 0$, and $C_{\varepsilon, R}$ is as follows:
\begin{figure}[h]
\begin{center}
\includegraphics[width=5cm,height=3cm]{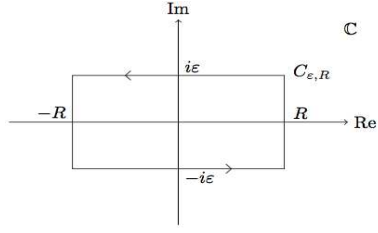}
\caption{The Contour $C_{\varepsilon, R}$}
\end{center}
\end{figure}

\noindent
By Proposition \ref{pro:resolvent of perturbed hamiltonian}, we have
\begin{align}
(z - h)^{-1} \Lvector{c \\ \psi} =& \frac{1}{ \eta(z) } \Lvector{ c \\ 0 } + \frac{\lambda}{ \eta(z) } \innpro{ g, (z - {h_{0,0}})^{-1} \psi } \Lvector{ 1 \\ 0 } + \Lvector{ 0 \\ (z - {h_{0,0}})^{-1} \psi } \nonumber\\
& + \frac{\lambda c}{ \eta(z) } \Lvector{ 0 \\ (z - {h_{0,0}})^{-1} g } + \frac{\lambda^2 \innpro{ g, (z - {h_{0,0}})^{-1} \psi }}{ \eta(z) } \Lvector{ 0 \\ (z - {h_{0,0}})^{-1} g }. \label{eq:resolvent hit vector}
\end{align}
The equation (\ref{eq:resolvent hit vector}), the definition of $\eta$, Lemma \ref{lmm:forms of terms of line_integrals}, and conditions {\rm (A)} and {\rm (B)} imply the following equation:
\begin{align}
& \frac{1}{2 \pi i} \lim_{R\nearrow \infty, \varepsilon \searrow 0} \int_{C_{\varepsilon,R}} \frac{e^{itz}}{ \eta(z) } dz = \lambda^2 \innpro{g, \frac{  e^{it h_{0,0}}}{ \abs{ \eta_-(h_{0,0}) }^2 }  g}, \\
& \frac{1}{2 \pi i} \lim_{R \nearrow \infty, \varepsilon \searrow 0} \int_{C_{\varepsilon,R}} \frac{\lambda e^{itz}}{ \eta(z) } \innpro{ g, (z - {h_{0,0}})^{-1} \psi } dz = \lambda \innpro{g, \frac{ e^{it h_{0,0}} }{ \eta_+(h_{0,0}) } \psi} + \lambda^3 \int_{\sigma_0} \frac{ e^{it\nu} \innpro{ g, (\nu - h_{0,0} - i0)^{-1} \psi } }{ \abs{\eta_-(\nu)}^2 } d\innpro{ g, E_0(\nu) g }, \\
& \frac{1}{2 \pi i} \lim_{R \nearrow \infty, \varepsilon \searrow 0} \int_{C_{\varepsilon,R}} \frac{\lambda c e^{itz}}{ \eta(z) } \innpro{ \xi, (z - {h_{0,0}})^{-1} g } dz \nonumber\\
=& \lambda c \innpro{\xi, \frac{ e^{it h_{0,0}} }{ \eta_-(h_{0,0}) } g} - \lambda^3 c \int_{\sigma_0} \frac{ e^{it\nu} \innpro{ \xi, (h_{0,0} - \nu - i0)^{-1} g } }{ \abs{\eta_-(\nu)}^2 } d\innpro{ g, E_0(\nu) g }, \\
& \frac{1}{2 \pi i} \lim_{R \nearrow \infty, \varepsilon \searrow 0} \int_{C_{\varepsilon, R}} \frac{\lambda^2 e^{it z}}{ \eta(z) }  \innpro{ g,(z - {h_{0,0}})^{-1} \psi} \innpro{\xi, (z - {h_{0,0}})^{-1} g} dz \nonumber\\
=& \lambda^2 \int_{\sigma_0} \frac{e^{it \nu}}{\eta_-(\nu)} \innpro{ g, (\nu - h_{0,0} - i 0)^{-1} \psi } d \innpro{ \xi, E_0(\nu) g } + \lambda^2 \int_{\sigma_0} \frac{e^{it \nu}}{\eta_+(\nu)} \innpro{ \xi,  (\nu - h_{0,0} + i 0)^{-1} g } d \innpro{g, E_0(\nu) \psi} \nonumber\\
& + \lambda^4 \int_{\sigma_0} \frac{ e^{it \nu} }{\abs{\eta_-(\nu)}^2}  \innpro{ g, (\nu - h_{0,0} - i 0)^{-1} \psi } \innpro{ \xi, (\nu - h_{0,0} + i 0)^{-1} g } d\innpro{ g, E_0(\nu) g }.
\end{align}
We conclude that
\begin{align*}
c(t) =& \lambda \innpro{g, \frac{ e^{it h_{0,0}} }{ \eta_+(h_{0,0}) } \psi} + \lambda^2 \int_{\sigma_0} \frac{ e^{it \nu} }{ \abs{ \eta_-(\nu) }^2 } \cbk{ c + \lambda \innpro{ g, (\nu - h_{0,0} - i0)^{-1} \psi } } d\innpro{g, E_0(\nu) g}, \\
\innpro{ \xi, \psi(t)} =& \innpro{\xi, e^{it h_{0,0}} \psi} + \lambda \int_{\sigma_0} \frac{ e^{it \nu} }{ \eta_-(\nu) } \cbk{ c + \lambda \innpro{g, ( \nu - h_{0,0} - i0 )^{-1} \psi} } d\innpro{\xi, E_0(\nu) g}  \nonumber\\
& - \lambda^2 \int_{\sigma_0} \frac{e^{it \nu}}{\eta_+(\nu)} \innpro{ \xi,  ( h_{0,0} -\nu - i 0)^{-1} g } d \innpro{g, E_0(\nu) \psi} \nonumber\\
& - \lambda^3 \int_{\sigma_0} \frac{ e^{it\nu} \innpro{ \xi, (h_{0,0} - \nu - i0)^{-1} g } }{ \abs{\eta_-(\nu)}^2 } \cbk{ c + \lambda \innpro{ g, (\nu - h_{0,0} - i 0)^{-1} \psi } } d\innpro{ g, E_0(\nu) g }.
\end{align*}
We have thus proved the theorem.\QED

\setcounter{theorem}{0}
\setcounter{equation}{0}
\section{NESS} \label{sec:NESS}
In this section, we give the initial state and an explicit formula of NESS.
In the present paper, we consider the case that $h_{0,k}$ is the multiplication operator of $\abs{p}^2/2$, $p \in \bbR^d$, on $L^2(\bbR^d)$, $d \geq 3$, or $h_{0,k} = \norm {A_{G_k}} \mathbbm{1} - A_{G_k}$, where $G_k$ are undirected connected graphs with bounded degree, with countable infinite vertices, with no loops, and with no multiple edges and $A_{G_k}$ is the adjacency operator of $G_k$ for each $k=1, \ldots, N$. 
We recall the definition of Perron--Frobenius weights  (PF weights, for short).

If $h_{0,k}$ is the multiplication operator of $\abs{p}^2/2$, then we define the PF weight $v_k$ by $v_k := \delta$, where $\delta$ is the Dirac's delta function.

Let $G$ be an undirected connected graph with bounded degrees, with countable infinite vertices, with no loops, and with no multiple edges.
We denote the set of vertices of $G$ by $VG$.

Fix a positive preserving operator $B$ acting on $\ell^2(VG)$.
The sequence $\{ v(x) \}_{x \in VG}$ is called a PF weight if it has positive entries and
\begin{align}
\sum_{y \in VG} B_{x y} v(y) = {\rm spr}(B) v(x), \quad x\in VG,
\end{align}
where ``{\rm spr}'' stands for spectral radius.
If such a vector belongs to $\ell^2(VG)$ it is a standard eigenvector for $B$.
For the adjacency operator $A_{G}$ of $G$, the existence of $v$ proved in \cite[Proposition 4.1]{FidaleoGuidoIsola}.
We regard a PF weight $v$ as a densely defined linear functional on $\ell^2(VG)$.
We define the domain $\calD(v)$ of $v$ by
\begin{align}
\calD(v) = \Set{ \psi \in \ell^2(VG) | \sum_{x \in VG} v(x) \abs{\psi(x)} < \infty },
\end{align}
where $\psi(x) = \innpro{\delta_x, \psi}$ and $\delta_x$ is the delta function such that $\delta_x(y) = 0$ for any $y \neq x$ and $\delta_x(x) = 1$. 
If $\psi \in \calD(v)$, we denote $\sum_{x \in VG} v(x) \psi(x)$ by $\innpro{v, \psi}$.
Note that PF weights for the adjacency operator of an infinite graph may not be unique.
(See e.g. \cite{FidaleoGuidoIsola}, \cite{Fidaleo}, and \cite{Pruitt}.)
When $v$ is a PF weight for $A_{G}$,  we say that $v$ is a PF weight for $\norm{A_G} \mathbbm{1} - A_G$.

For an operator $A$ on a Hilbert space, we denote the domain of $A$ by $\calD(A)$.
For each $k=1, \ldots, N$, we denote the inverse temperature and the chemical potential of $k$-th reservoir by $\beta_k > 0$ and $\mu_k \leq 0$, respectively.
Let $v_k$ be PF weights for $h_{0,k}$, $k=1, \ldots, N$.
We set $\calD(v) = \bigoplus_{k=1}^N \calD(v_k)$ and $(\psi_k) = {}^t(0, 0, \ldots, 0, \psi_k, 0, \ldots, 0)$ for $\psi_k \in \fraK_k$, i.e. $(\psi_k)$ means the $k$-th component is equal to $\psi_k$ and the others are equal to zero. 
Suppose $\psi_k \in \calD(v_k) \cap \calD((e^{\beta_k (h_{0,k} - \mu_k)} - \mathbbm{1})^{-1/2})$.
We consider the initial state  $\omega_0$ of the $k$-th reservoir given by
\begin{align}
\omega_0(W( (\psi_k) )) = \exp\rbk{ -\frac{1}{2} \innpro{ \psi_k, \rbk{\calN_k(h_{0,k}) + 1/2} \psi_k } } e^{i \Theta_k(\innpro{v_k, \psi_k})}, \label{eq:initial state}
\end{align}
where 
\begin{align}
\calN_k(x) = (e^{\beta_k (x - \mu_k) } - 1)^{-1}
\end{align}
and $\Theta_k$ is real valued real linear functional on $\bbC$.
Detail forms of $\Theta_k(\innpro{v_k, \cdot})$ is given in Section \ref{sec:examples}.
We assume that if $\mu_k < 0$, then $\Theta_k \equiv 0$.
To obtain an explicit formula of NESS, we assume that the following conditions for initial states and form factors:

\begin{description}
\item[ {\rm (C)} ] The initial state $\omega_0$ satisfies
\begin{align}
\abs{ \omega_0(a^{\natural_1} a^{\natural_2} \cdots a^{\natural_n})  } \leq n! K_n, \quad n \in \bbN,
\end{align}
where $a^{\natural_j} = a ({}^t(1, 0))$ or $a^\dagger ({}^t(1, 0))$ and $K_n(>0)$ satisfies $\lim_{n \to \infty} K_{n + 1}/ K_n = 0$.
\item[{\rm (D)}] The form factors $g_k$ are in $\calD(v_k) \cap \calD((e^{\beta_k h_{0,k} } - \mathbbm{1})^{-1/2})$, $k=1, \ldots, N$.
\end{description}

\begin{lmm} \label{lmm:perturbed domain}
If the form factors $g_k$ and vectors $\psi_k$ belong to $\calD((e^{\beta_k h_{0,k}} - \mathbbm{1})^{-1/2})$, then $P_k e^{ith} \, {}^t(c, \psi_1, \ldots, \psi_N)$ is in $\calD((e^{\beta_k h_{0,k}} - \mathbbm{1})^{-1/2})$ for any $t \in \bbR$ and $c \in \bbC$, $k = 1, \ldots, N$, where $P_k$ is the projection from $\calK$ onto $\fraK_k$.
\end{lmm}

\noindent
{\bf Proof.} 
For simplicity, we assume that $t > 0$.
For any $c \in \bbC$ and $\psi_k \in \calD((e^{\beta_k h_{0,k}} - \mathbbm{1})^{-1/2})$, $k = 1, \ldots, N$, we have that
\begin{align}
e^{ith} \Lvector{c \\ \psi} = \sum_{n \geq 0} \lambda^n i^n \int_0^t dt_n \cdots \int_0^{t_2} dt_1 \alpha_{t_1}^0(V) \alpha_{t_2}^0(V) \cdots \alpha_{t_n}^0(V) e^{ith_0} \Lvector{c \\ \psi}
\end{align}
by Dyson series expansion, where $\alpha^0_t(V) = e^{ith_0} V e^{-ith_0}$, $V$ is the operator defined in (\ref{eq:coupled hamiltonian}), and $\psi = {}^t(\psi_1, \ldots, \psi_N)$.
For $n \geq 1$, we obtain
\begin{align}
\alpha_{t_1}^0(V) \alpha_{t_2}^0(V) \cdots \alpha_{t_n}^0(V) e^{ith_0} \Lvector{c \\ \psi} =& \innpro{ \Lvector{1 \\ 0}, e^{-it_1h_0} \alpha_{t_2}^0(V) \cdots \alpha_{t_n}^0(V) e^{ith_0} \Lvector{c \\ \psi}} e^{it_1h_0} \Lvector{0 \\ g} \nonumber\\
& + \innpro{\Lvector{0 \\ g}, e^{-it_1h_0} \alpha_{t_2}^0(V) \cdots \alpha_{t_n}^0(V) e^{ith_0} \Lvector{c \\ \psi}} e^{it_1h_0} \Lvector{1 \\ 0} 
\end{align}
and 
\begin{align}
\norm{ (e^{\beta_k h_{0,k}} - \mathbbm{1})^{-1/2} P_k \alpha_{t_1}^0(V) \alpha_{t_2}^0(V) \cdots \alpha_{t_n}^0(V) e^{ith_0} \Lvector{c \\ \psi} } \leq \norm{V}^{n-1} \norm{\Lvector{c \\ \psi}} \norm{(e^{\beta_k h_{0,k}} - \mathbbm{1})^{-1/2} g_k}.
\end{align}
It follows that
\begin{align}
\norm{(e^{\beta_k h_{0,k}} - \mathbbm{1})^{-1/2} P_k e^{ith} \Lvector{c \\ \psi}} &\leq \norm{ (e^{\beta_k h_{0,k}} - \mathbbm{1})^{-1/2} \psi_k } +  \norm{(e^{\beta_k h_{0,k}} - \mathbbm{1})^{-1/2} g_k} \norm{\Lvector{c \\ \psi}} \sum_{n \geq 1} \lambda^n  \norm{V}^{n-1} \frac{t^n}{n!} \nonumber\\
&= \norm{ (e^{\beta_k h_{0,k}} - \mathbbm{1})^{-1/2} \psi_k } + \frac{ e^{\lambda t \norm{V}} - 1 }{\norm{V}} \norm{(e^{\beta_k h_{0,k}} - \mathbbm{1})^{-1/2} g_k} \norm{\Lvector{c \\ \psi}} < \infty.
\end{align}
This completes the  proof of Lemma \ref{lmm:perturbed domain}. \QED

\begin{rmk} \label{rmk:remark of domains}
Note that $\calD((e^{\beta_k h_{0,k}} - \mathbbm{1})^{-1/2}) = \calD((h_{0,k})^{-1/2})$. 
(cf. Paragraphs before {\rm \cite[Theorem 4.5]{Fidaleo}}.)
In fact, we consider the continuous function
\begin{align}
q(x) = \left\{
\begin{array}{cc}
-\frac{1}{2} & x=0 \\
(e^{x} - 1)^{-1} - x^{-1}  & x > 0 
\end{array}
\right. .
\end{align}
The function $q$ is bounded on $[0, \infty)$ and $(e^x - 1)^{-1} = q(x) + x^{-1}$, $x \in (0,\infty)$.
For any $\varepsilon > 0$, the following equation holds:
\begin{align}
\innpro{\psi, (e^{\beta_k (h_{0,k} + \varepsilon)} - \mathbbm{1})^{-1} \psi} = \innpro{\psi, q(\beta_k (h_{0,k} + \varepsilon)) \psi} + \innpro{\psi, (\beta_k(h_{0,k} + \varepsilon))^{-1} \psi}. \label{eq:inverse domain}
\end{align}
If $\psi \in \calD(e^{\beta_k h_{0,k}} - \mathbbm{1})^{-1/2}$, then 
\begin{align}
\lim_{\varepsilon \downarrow 0} \innpro{\psi, (e^{\beta_k (h_{0,k} + \varepsilon)} - \mathbbm{1})^{-1} \psi} < \infty.
\end{align}
By the boundedness of $q$, $\lim_{\varepsilon \downarrow 0} \innpro{\psi, (\beta_k (h_{0,k}+ \varepsilon))^{-1} \psi} < \infty$.
Thus, $\calD((e^{\beta_k h_{0,k}} - \mathbbm{1})^{-1/2}) \subset \calD((h_{0,k})^{-1/2})$.
By similar discussion, we obtain $\calD((h_{0,k})^{-1/2}) \subset \calD((e^{\beta h_{0,k}} - \mathbbm{1})^{-1/2})$.
As a consequence, we have that $\calD((e^{\beta_k h_{0,k}} - \mathbbm{1})^{-1/2}) = \calD((h_{0,k})^{-1/2})$.
\end{rmk}

We define the upper half-plane $\bbC_+$ on $\bbC$ by $\bbC_+ := \set{ z \in \bbC | {\rm Im} z > 0 }$ and the Hardy space $\bbH^\infty(\bbC_+)$ on the upper half-plane defined by
\begin{align}
\bbH^\infty(\bbC_+) := \Set{ f : \text{ holomorphic on } \bbC_+ | \norm{f}_{\infty} :=  \sup_{z \in \bbC_+} \abs{ f(z) } < \infty }.
\end{align}
We denote the Hardy space over the lower half-plane $\bbC_-$ by $\bbH^\infty(\bbC_-)$.

\begin{theorem} \label{theorem:general Non-equilibrium steady states}
Under conditions {\rm (A)} $\sim$ {\rm (D)}, we have that
\begin{align}
\lim_{t \to + \infty} \omega_0 \circ \alpha_t(W(f)) = \exp \cbk{ - \frac{1}{2} S(f) + i\Lambda(f) } =: \omega_+(W(f)) 
\end{align}
for any $c \in \bbC$, any $\psi \in \frakk := \frah(g) \cap \calD(v) \cap (\bigoplus_{k=1}^N \calD((e^{\beta_k h_{0,k} } - \mathbbm{1})^{-1/2}) )$, and $f ={}^t(c, \psi)$, where
\begin{align}
S(f) = \sum_{l = 1}^N \innpro{\varphi_l(f), (\calN_l(h_{0,l}) + 1/2) \varphi_l(f)}, \quad \Lambda(f) =& \sum_{l=1}^N \Theta_l(\innpro{v_l, \varphi_l(f)}),
\end{align}
and $\innpro{v_l, \varphi_l(f)}$ is defined by
\begin{align}
\innpro{v_l, \varphi_l(f)} := \innpro{v_l, \psi_l} + \frac{\lambda c \innpro{v_l, g_l}}{\eta(0)} + \frac{\lambda^2 }{\eta(0)}  \innpro{v_l, g_l} \innpro{ g, (h_{0,0})^{-1} \psi }. \label{eq:def of linear form of PF weights}
\end{align}
\end{theorem}

\begin{rmk}
The above theorem implies that NESS $\omega_+$ exists uniquely and has the form of
\begin{align}
\omega_+(\Psi(f)) =& \pi^{3/2} \sum_{l=1}^N \Theta_l(\innpro{v_l, \varphi_l(f)}), \label{eq:linear part of quasi-free state}\\
\omega_+(\Psi(f)^2) - \omega_+(\Psi(f))^2 =& \sum_{l=1}^N \innpro{\varphi_l(f), (\calN_l(h_{0,l}) + 1/2) \varphi_l(f)}. \label{eq:quadratic part of quasi-free state}
\end{align}
\end{rmk}

\noindent
{\bf Proof of Theorem \ref{theorem:general Non-equilibrium steady states}.}
For a vector $f \in \calK$, we denote the scalar part and $\fraK_k$-part of $f$ by $f_0$ and $f_k$, $k=1, \ldots, N$, respectively.
By (\ref{eq:initial state}), for $c \in \bbC$, $\psi \in \frakk$, and $f = {^t(c, \psi)}$, we have that
\begin{align}
\omega_0 \circ \alpha_t( W(f)) = \omega_0(W( (e^{ith} f)_0 )) \prod_{k=1}^N \omega_0(W( (e^{ith} f)_k )).
\end{align}

First, we consider the limit of $\omega_0(W( (e^{ith} f)_0 ))$.
Condition {\rm (C)} and Theorem \ref{theorem:general Time evolution} imply
\begin{align}
\abs{ \omega_0( \cbk{ \Phi((e^{ith} f)_0) }^m) } \leq m! (2 \abs{c(t)})^m K_m.
\end{align}
Since $1/\eta_- \in L^\infty(\bbR)$, 
\begin{align}
\frac{1}{ \eta_+(\nu) } \frac{d \innpro{ g, E_0 (\nu) \psi }}{ d \nu}, \quad \frac{F(\nu; f)}{\abs{\eta_-(\nu)}^2} \frac{d \innpro{ g, E_0(\nu) g } }{d \nu} \in L^1(\bbR).
\end{align}
Thus, $c(t)$ is bounded independent of $t$.
A theorem of Riemann--Lebesgue (see e.g. \cite{Titchmarsh}) implies $c(t) \to 0$ as $t \to \infty$.
We obtain
\begin{align}
& \abs{\omega_0(\exp(i \Phi( (e^{ith} f)_0 ) )) - 1} \leq \sum_{m  = 1}^\infty \frac{1}{m!} \abs{ \omega_0(\cbk{\Phi(e^{ith} f)_0 }^m) } \leq \sum_{m = 1}^\infty (2 \abs{c(t)})^m K_m \nonumber\\
=& \frac{\abs{c(t)}}{C} \sum_{m = 1}^\infty (2 C)^m K_m \to 0, \quad  (t \to \infty),
\end{align}
where $C := \sup_{t \in \bbR}\abs{c(t)}$.

Next, we consider the quadratic part of $\log \omega_0(W((e^{ith} f)_l))$.
For $\varepsilon \in (0, \pi/2)$, we put 
\begin{align}
\psi_{\varepsilon, l}(t) := e^{ith_{0, l}} \varphi_l(f) - \lambda^2 \int_{\sigma_0} \frac{e^{it \nu}}{ \eta_+(\nu)} (h_{0,l} - \nu - i \varepsilon)^{-1} g_l d \innpro{g, E_0(\nu) \varphi(f)}, \label{eq:vector integral}
\end{align}
where the convergence of vector valued integral of (\ref{eq:vector integral}) is in the strong operator topology.
Note that $(h_{0,l} - \nu - i \varepsilon)^{-1}$ is bounded.
We have that
\begin{align}
& \innpro{ \psi_{\varepsilon, l}(t), \cbk{ \calN_{l}(h_{0,l}) + 1/2 } \psi_{\varepsilon, l}(t) } \nonumber\\
=& \innpro{ \varphi_l(f), \cbk{ \calN_{l}(h_{0,l}) + 1/2 } \varphi_l(f) } \nonumber\\
&- \lambda^2 {\rm Re} \cbk{ \int_{\sigma_0} \frac{e^{it \nu}}{\eta_+(\nu)} \innpro{ e^{ith_{0,l} } \varphi_l(f), \cbk{ \calN_l(h_{0,l}) + 1/2 } (h_{0,l} - \nu - i\varepsilon)^{-1} g_l } d \innpro{g, E_0(\nu) \varphi(f)} } \nonumber\\
& + \lambda^4 \int_{\sigma_0} \int_{\sigma_0} \frac{e^{it(\nu - \nu^\prime)}}{ \eta_-(\nu^\prime) \eta_+(\nu) } \innpro{ (h_{0,l} - \nu^\prime - i\varepsilon )^{-1} g_l, \cbk{ \calN_l(h_{0,l}) + 1/2 }(h_{0,l} - \nu - i\varepsilon)^{-1} g_l } d \innpro{g, E_0(\nu) \varphi(f)} d \innpro{\varphi(f), E_0(\nu^\prime) g}. \label{eq:integration A}
\end{align}
The second term on the right side in (\ref{eq:integration A}) has the following form:
\begin{align}
& \int_{\sigma_0} \frac{e^{it \nu}}{\eta_+(\nu)} \innpro{ e^{ith_{0,l} } \varphi_l(f), \cbk{ \calN_l(h_{0,l}) + 1/2 } (h_{0,l} - \nu - i\varepsilon)^{-1} g_l } d \innpro{g, E_0(\nu) \varphi(f)} \nonumber\\
=& \int_{\sigma_0} \frac{1}{\eta_+(\nu)} \int_{\sigma_l} \frac{e^{it(\nu - \nu^\prime)}}{\nu^\prime - \nu - i\varepsilon} \cbk{ \calN_l(\nu^\prime) + 1/2 } d\innpro{ \varphi_l(f),  E_l(\nu^\prime) g_l } d \innpro{g, E_0(\nu) \varphi(f)} \nonumber\\
=& i e^{t \varepsilon} \int_{\sigma_0} \frac{1}{\eta_+(\nu)} \int_{\sigma_l} \int_0^\infty e^{i(t+s)(\nu - \nu^\prime + i\varepsilon)} ds \cbk{ \calN_l(\nu^\prime) + 1/2 } d\innpro{ \varphi_l(f),  E_l(\nu^\prime) g_l } d \innpro{g, E_0(\nu) \varphi(f)} \nonumber\\
=& i e^{t \varepsilon} \int_t^\infty e^{- s \varepsilon} \int_{\sigma_0} \frac{e^{is \nu} }{\eta_+(\nu)} d \innpro{g, E_0(\nu) \varphi(f)} \int_{\sigma_l} e^{- is \nu^\prime} \cbk{ \calN_l(\nu^\prime) + 1/2 } d\innpro{ \varphi_l(f),  E_l(\nu^\prime) g_l } ds.
\end{align}
Since $g, \psi \in \bigoplus_{k=1}^N \calD((e^{\beta_k h_{0,k}} - \mathbbm{1})^{-1/2})$, we have that $g, \psi \in  \bigoplus_{k=1}^N \calD((h_{0,k})^{-1/2})$ by Remark \ref{rmk:remark of domains}.
Thus, we obtain $\varphi_l(f) \in \calD((h_{0,l})^{-1/2})$ for any $l = 1,\ldots, N$ and 
\begin{align}
 \calN_l(\nu^\prime) \frac{d\innpro{ \varphi_l(f),  E_l(\nu^\prime) g_l }}{d \nu^\prime} = \nu^\prime \calN_l(\nu^\prime) \frac{d\innpro{ (h_{0,l})^{-1/2} \varphi_l(f),  E_l(\nu^\prime) (h_{0,l})^{-1/2} g_l }}{d \nu^\prime}. \label{eq:www}
\end{align}
The above equation (\ref{eq:www}) and conditions {\rm (A)}, {\rm (B)}, and {\rm (D)} imply that the following functions
\begin{align}
\frac{1 }{\eta_+(\nu)} \frac{d \innpro{g, E_0(\nu) \varphi(f)}}{d\nu}, \cbk{ \calN_l(\nu^\prime) + 1/2 } \frac{d\innpro{ \varphi_l(f),  E_l(\nu^\prime) g_l }}{d \nu^\prime} \in L^1(\bbR) \cap L^2(\bbR). \label{eq:bddness}
\end{align}
Thus, there exist functions $w_1, w_2 \in L^2(\bbR)$ such that 
\begin{align}
& \int_t^\infty e^{- s \varepsilon} \int_{\sigma_0} \frac{e^{is \nu} }{\eta_+(\nu)} d \innpro{g, E_0(\nu) \varphi(f)} \int_{\sigma_l} e^{- is \nu^\prime} \cbk{ \calN_l(\nu^\prime) + 1/2 } d\innpro{ \varphi_l(f),  E_l(\nu^\prime) g_l } ds \nonumber\\
=& \int_{t}^\infty e^{- s  \varepsilon} \overline{w_1(s)} w_2(s) ds \label{eq:int1}
\end{align}
by Plancherel theorem.
The integral (\ref{eq:int1}) is absolutely convergent independent of $\varepsilon$ and $t$.

We consider the third term on the right side in (\ref{eq:integration A}).
Since $\calN_l(x) + 1/2$ is in $L^1(\bbR, d\innpro{ g_l, E_l(x) g_l })$, the third term on the right side in (\ref{eq:integration A}) has the form of
\begin{align}
&\lambda^4 \int_{\sigma_0} \int_{\sigma_0} \frac{e^{it(\nu - \nu^\prime)}}{ \eta_-(\nu^\prime) \eta_+(\nu) } \innpro{ (h_{0,l} - \nu^\prime - i\varepsilon )^{-1} g_l,\cbk{ \calN_l(h_{0,l}) + 1/2 } (h_{0,l} - \nu - i\varepsilon)^{-1} g_l } d \innpro{g, E_0(\nu) \varphi(f)} d \innpro{\varphi(f), E_0(\nu^\prime) g} \nonumber\\
=& \lambda^4 \lim_{\delta \searrow 0}  \int_{\sigma_0} \int_{\sigma_0} \frac{e^{it(\nu - \nu^\prime)}}{ \eta_-(\nu^\prime) \eta_+(\nu) } \int_{\sigma_l} \frac{ \calN_l(x + \delta) + 1/2 }{(x - \nu^\prime + i \varepsilon)(x - \nu - i\varepsilon)} d \innpro{g_l, E_l(x) g_l}  d \innpro{g, E_0(\nu) \varphi(f)} d \innpro{\varphi(f), E_0(\nu^\prime) g}. \label{eq:getahaki}
\end{align} 
We define the holomorphic function $u_\delta$ in $\bbC_+$ by
\begin{align}
u_\delta(z) (= u_\delta(x + i y)) := \frac{1}{\pi} \cbk{ \calN_l(z + \delta) + 1/2} \int_{\bbR} \frac{y}{ (x - w)^2 + y^2 } \frac{d \innpro{ g_l, E_l(w) g_l }}{d w} dw.
\end{align}
Note that the support of $d\innpro{ g_l, E_l(x) g_l} / dx$ is contained in $[0, \infty)$, $d\innpro{ g_l, E_0(w) g_l} / dw$ is in $L^\infty(\bbR)$ by condition {\rm (A)}, and $\calN_l(z + \delta)$ is analytic and bounded in $D := \set{ z \in \bbC | {\rm Im} z > 0, -\delta^\prime < {\rm Re} z < \infty }$, where $0 < \delta^\prime < \delta$.
Thus, for $z \in \bbC_+$ with ${\rm Re}z < 0$, $u_\delta(z) = 0$, $u_\delta \in \bbH^\infty(\bbC_+)$, and $u_\delta(x + i y)$ converges to $\cbk{\calN_l(x + \delta)+1/2} d\innpro{ g_l, E_0(x) g_l} /dx$ as $y \searrow 0$ in $L^\infty(\bbR)$ by \cite[Theorem 3.13.]{Jaksic06}.
For any $R > \norm{h_{0,0}}$, we obtain the following equation:
\begin{align}
&\int_{\sigma_l} \frac{ \calN_l(x + \delta) + 1/2 }{(x - \nu^\prime + i \varepsilon)(x - \nu - i\varepsilon)} d \innpro{g_l, E_l(x) g_l} = \lim_{R \to \infty, \delta^\prime \searrow 0} \int_{-R}^R \frac{u_\delta(x + \delta^\prime)}{(x - \nu^\prime + i\varepsilon)(x - \nu - i\varepsilon)} dx \nonumber\\
=& \lim_{R \to \infty, \delta^\prime \searrow 0} \int_{\gamma_{R, \delta^\prime}}  \frac{u_\delta(z + \delta^\prime)}{(z - \nu^\prime + i\varepsilon)(z - \nu - i\varepsilon)} dz = 2 \pi i \frac{u_\delta(\nu + i \varepsilon)}{\nu - \nu^\prime + 2 i \varepsilon},
\end{align}
where $\gamma_{R, \delta^\prime}$ is the contour from $[-\delta^\prime, R]$ through $[R, R + iR]$ and $[R+iR, -\delta^\prime + iR]$ to $[-\delta + iR, -\delta^\prime]$.
By (\ref{eq:bddness}), (\ref{eq:getahaki}), and $\varepsilon \in (0, \pi/2)$, the last term on the right side in (\ref{eq:integration A}) has the form of
\begin{align}
& \lambda^4 \int_{\sigma_0} \int_{\sigma_0} \frac{e^{it(\nu - \nu^\prime)}}{ \eta_-(\nu^\prime) \eta_+(\nu) } \innpro{ (h_{0,l} - \nu^\prime - i\varepsilon )^{-1} g_l, \rbk{\calN_l(h_{0,l}) + 1/2} (h_{0,l} - \nu - i\varepsilon)^{-1} g_l } d \innpro{g, E_0(\nu) \varphi(f)} d \innpro{\varphi(f), E_0(\nu^\prime) g} \nonumber\\
=& 2 \pi i \lambda^4 \lim_{\delta \searrow 0} \int_{\sigma_0} \int_{\sigma_0} \frac{e^{it(\nu - \nu^\prime)}}{ \eta_-(\nu^\prime) \eta_+(\nu) } \frac{u_\delta(\nu + i\varepsilon)}{\nu - \nu^\prime + 2 i\varepsilon} d \innpro{g, E_0(\nu) \varphi(f)} d \innpro{\varphi(f), E_0(\nu^\prime) g} \nonumber\\
=& 2 \pi \lambda^4 e^{2 t \varepsilon} \int_{\sigma_0} \int_{\sigma_0}  \frac{ u(\nu + i\varepsilon) }{ \eta_-(\nu^\prime) \eta_+(\nu) }  \int_t^\infty e^{i s(\nu - \nu^\prime + 2 i\varepsilon)} ds d \innpro{g, E_0(\nu) \varphi(f)} d \innpro{\varphi(f), E_0(\nu^\prime) g} \nonumber\\
=& 2 \pi \lambda^4  e^{2 t \varepsilon} \int_{t}^\infty e^{- 2 s  \varepsilon} \int_{\sigma_0}  \frac{ e^{is \nu} u(\nu + i \varepsilon) }{ \eta_+(\nu) } d \innpro{g, E_0(\nu) \varphi(f)} \int_{\sigma_0} \frac{ e^{- is \nu^\prime}}{\eta_-(\nu^\prime)} d \innpro{\varphi(f), E_0(\nu^\prime) g} ds,
\end{align}
where $u(\nu + i\varepsilon) = \lim_{\delta \searrow 0} u_\delta(\nu + i\varepsilon)$.
By (\ref{eq:bddness}), there exist functions $w_3, w_4 \in L^2(\bbR)$ such that 
\begin{align}
& \int_{t}^\infty e^{- 2 s  \varepsilon} \int_{\sigma_0}  \frac{ e^{is \nu} u(\nu + i \varepsilon)}{ \eta_+(\nu) } d \innpro{g, E_0(\nu) \varphi(f)} \int_{\sigma_0} \frac{ e^{- is \nu^\prime}}{\eta_-(\nu^\prime)} d \innpro{\varphi(f), E_0(\nu^\prime) g} ds \nonumber\\
=& \int_{t}^\infty e^{- 2 s  \varepsilon} \overline{w_3(s)} w_4(s) ds \label{eq:int2}
\end{align}
by Plancherel theorem. The integral (\ref{eq:int2}) is absolutely convergent independent of $\varepsilon$ and $t$.
Therefore, we have that
\begin{align}
& \innpro{ \psi_l(t), \cbk{ \calN_l(h_{0,l}) + 1/2 } \psi_l(t) } = \lim_{\varepsilon \searrow 0} \innpro{ \psi_{\varepsilon, l}(t), \cbk{ \calN_l(h_{0,l}) + 1/2 } \psi_{\varepsilon,l}(t) } \nonumber\\ 
=& \innpro{ \varphi_l(f), \cbk{ \calN_{l}(h_{0,l}) + 1/2 } \varphi_l(f) } - \lambda^2 {\rm Re} \cbk{ i \int_{t}^\infty \overline{w_1(s)} w_2(s) ds } + 2 \pi \lambda^4 \int_{t}^\infty \overline{w_3(s)} w_4(s) ds.
\end{align}
Thus, we obtain
\begin{align}
\lim_{t \to \infty} \innpro{ \psi_l(t), \cbk{ \calN_l(h_{0,l}) + 1/2 } \psi_l(t) } = \innpro{ \varphi_l(f), \cbk{ \calN_{l}(h_{0,l}) + 1/2 } \varphi_l(f) }.
\end{align}

Finally, we discuss the term $\Theta_l\left(\innpro{v_l, (e^{ith}f)_l }\right)$ in (\ref{eq:initial state}).
If $h_{0,k}$ are the multiplication operator of $\abs{p}^2/2$, then we obtain the statement by \cite[Theorem 3.1]{MatsuiTasaki}.
Thus, we consider the case that $h_{0,k}$ are the adjacency operators of graphs.
For $z \in \bbC \backslash \sigma_k$ and $\xi \in \frakk_k := P_k \frakk$, we obtain
\begin{align}
\abs{ \innpro{ \delta_x, (z - h_{0,k})^{-1} \xi } } = \abs{ \int_{\sigma_k} (z - \nu)^{-1} d \innpro{ \delta_x, E_k(\nu) \xi } } \leq \sup_{\nu \in \sigma_k} \abs{z - \nu}^{-1} \int_{\sigma_k} d \abs{ \innpro{\delta_x, E_0(\nu) \xi} } = \sup_{\nu \in \sigma_k} \abs{z - \nu}^{-1} \abs{ \innpro{\delta_x, \xi} }.
\end{align}
Since $\xi \in \frakk_k \subset \calD(v_k)$, $(z - h_{0,k})^{-1}\xi \in \calD(v_k)$ for any $z \in \bbC \backslash \sigma_k$.
It follows that
\begin{align}
\innpro{ v_k, (z - h_{0,k})^{-1} \xi } = z^{-1} \innpro{ v_k, (z - h_{0,k}) (z - h_{0,k})^{-1} \xi } = z^{-1} \innpro{ v_k, \xi }.
\end{align}
By condition (D), the above discussion, and a theorem of Riemann--Lebesgue, we obtain
\begin{align}
\lim_{t \to \infty} \Lambda_l(e^{ith} f) =& \lim_{t \to \infty} \left[ \Theta_l\rbk{ \frac{\innpro{v, \psi_l} }{2 \pi i} \lim_{R \nearrow 0, \varepsilon \searrow 0} \int_{C_{R, \varepsilon}} \frac{e^{itz}}{z} dz } +  \Theta_l\rbk{ \frac{\lambda c \innpro{v_l, g_l}}{2 \pi i} \lim_{R \nearrow 0, \varepsilon \searrow 0}  \int_{C_{R, \varepsilon}} \frac{e^{itz}}{\eta(z) z} dz} \right. \nonumber\\
& \left. + \Theta_l \rbk{ \frac{ \lambda^2 \innpro{v_l, g_l}}{2 \pi i} \lim_{R \nearrow 0, \varepsilon \searrow 0}  \int_{C_{R, \varepsilon}} \frac{e^{itz} \innpro{ g, (z - h_{0,0})^{-1} \psi }}{\eta(z) z} dz} \right] = \Theta_l(\innpro{v_l, \varphi_l(f)}). \label{eq:form of linear part}
\end{align} 
We have completed the proof of Theorem \ref{theorem:general Non-equilibrium steady states}.\QED

\setcounter{theorem}{0}
\setcounter{equation}{0}
\section{Currents and Entropy Production Rate} \label{sec:general currents and entropy production rate}
We set $(c) = {}^t(c, 0, \ldots, 0)$, $c \in \bbC$.
Following ideas of W. Aschbacher et al. \cite{Aschbacher06} and \cite{AschbacherJaksic}, V. Jak\v{s}i\'{c} et al. \cite{Jaksic01}, for any $l = 1, \ldots, N$, currents $J_l$ and $E_l$ from $l$-th reservoir to the system is formally defined by
\begin{gather}
J_l = i \lambda a((1)) a^\dagger((g_l)) - i \lambda a ((g_l)) a^\dagger((1)), \label{eq:obs of charge} \\
E_l = i \lambda a((1)) a^\dagger((h_{0,l} g_l)) - i \lambda a((h_{0,l} g_l)) a^\dagger((1)), \label{eq:obs of heat}
\end{gather}
which are given by the following formal equations:
\begin{align}
\left. - \frac{d}{dt} \tau_t( d\Gamma(P_0) ) \right|_{t = 0} = \sum_{l=1}^N J_l, \quad \left. - \frac{d}{dt} \tau_t(d \Gamma(P_0 h_0) ) \right|_{t = 0} = \sum_{l=1}^N E_l,
\end{align}
where $P_0$ is the projection from $\calK$ onto $0 \oplus (\bigoplus_{l=1}^n \fraK_l)$.
Moreover, we define the mean entropy production rate by
\begin{align}
Ep(\omega_+) = \omega_+(\sigma),
\end{align}
where 
\begin{align}
\sigma = - \sum_{k=1}^N \beta_l(E_l - \mu_l J_l).
\end{align}

\begin{cor} \label{cor:general Currents}
Currents at NESS are given by the following form:
\begin{align}
\omega_+(J_l) =& 2 \pi \lambda^4 \sum_{k=1}^N \int_{\sigma_l} \frac{1}{ \abs{\eta_-(\nu)}^2 } \rbk{ \calN_l(\nu) - \calN_k(\nu) } \frac{d \innpro{g_k, E_k(\nu) g_k} }{d\nu} d\innpro{g_l, E_l(\nu) g_l} \nonumber\\
&  + \frac{\pi^3 \lambda^2}{\eta(0)} \sum_{k = 1}^N \cbk{ \Theta_k \rbk{ \innpro{v_k, g_k} } \Theta_l \rbk{ i \innpro{v_l, g_l} } - \Theta_k \rbk{   i \innpro{v_k, g_k} } \Theta_l \rbk{ \innpro{v_l, g_l} } }, \label{eq:general charge current} \\
\omega_+(E_l) =& 2 \pi \lambda^4 \sum_{k=1}^N \int_{\sigma_l} \frac{\nu}{ \abs{\eta_-(\nu)}^2 } \rbk{ \calN_l(\nu) - \calN_k(\nu) } \frac{d \innpro{g_k, E_k(\nu) g_k} }{d\nu} d\innpro{g_l, E_l(\nu) g_l}. \label{eq:general heat current}
\end{align}
\end{cor}

\noindent
{\bf Proof.}
For any $l = 1, \ldots, N$, $c \in \bbC$, and $\xi_l \in \frakk_l$, we have
\begin{align}
a((c)) a^\dagger ((\xi_l)) - a^\dagger((c)) a((\xi_l)) = i \Psi((ic))  \Psi((\xi_l)) - i \Psi((c)) \Psi((i\xi_l)).
\end{align}
Since $\sbk{  \Psi((c)), \Psi((\xi_l))   } = 0$, we obtain
\begin{align}
& 4 \omega_+( \Psi((c)) \Psi((\xi_l))) = \omega_+( \cbk{ \Psi((c) + (\xi_l)) }^2 ) - \omega_+ ( \cbk{ \Psi ((c) - (\xi_l)) }^2 )  \nonumber \\
=& \omega_+( \cbk{ \Psi((c) + (\xi_l)) }^2 ) - \omega_+ ( \Psi ((c) + (\xi_l)) )^2 - \omega_+( \cbk{ \Psi((c) - (\xi_l)) }^2 ) + \omega_+ ( \Psi ((c) - (\xi_l)) )^2  \nonumber\\
& + 4 \omega_+( \Psi ((c) ) ) \omega_+( \Psi ((\xi_l)) ). \label{eq:quasi-free form}
\end{align}
By Theorem \ref{theorem:general Non-equilibrium steady states} and (\ref{eq:quasi-free form}), it follows that
\begin{align}
& \omega_+(\Psi((1)) \Psi((i \xi_l))) - \omega_+(\Psi((i)) \Psi( (\xi_l))) \nonumber\\
=& 2 \lambda {\rm Im} \innpro{\xi_l, \cbk{ (\calN_l(h_{0,l}) + 1/2) / \eta_-(h_{0,l}) } g_l} + 2 \lambda^2 \sum_{k = 1}^N {\rm Im} \innpro{F(h_{0,k}, (\xi_l)) g_k, \cbk{ (\calN_k(h_{0,k}) + 1/2) / \abs{\eta_-(h_{0,k}}^2) } g_k} \nonumber\\
&  + \pi^3 \left\{ \rbk{ \sum_{k = 1}^N \Theta_k ( \innpro{v_k, \varphi_k( (1) ) } ) } \rbk{ \sum_{k=1}^N \Theta_k( \innpro{v_k, \varphi_k(  (i\xi_l) ) } ) } - \rbk{ \sum_{k=1}^N \Theta_k( \innpro{v_k, \varphi_k( (i) ) } )  }\rbk{ \sum_{k=1}^N \Theta_k( \innpro{v_k, \varphi_k( (\xi_l) ) } ) } \right\} \label{eq:Currents cal1}
\end{align}
by linearity of $\varphi_k(f)$ in $f$.
Note that the element $\xi_l$ is equal to $g_l$ or $h_{0,l} g_l$.
Thus the first term of (\ref{eq:Currents cal1}) has the form of
\begin{align}
& {\rm Im} \innpro{\xi_l, \cbk{ (\calN_l(h_{0,l}) + 1/2) / \eta_-(h_{0,l}) } g_l} \nonumber\\
=& \lambda^2 \pi \sum_{k=1}^N \int_{\sigma_l} \frac{1}{ \abs{\eta_-(\nu)}^2 } \rbk{ \calN_l(\nu) + \frac{1}{2} } \frac{d \innpro{g_k, E_k(\nu) g_k} }{d\nu} d\innpro{\xi_l, E_l(\nu) g_l}. \label{eq:1st term}
\end{align}

If $\xi_l = g_l$ or $h_{0, l} g_l$, then 
\begin{align}
{\rm Im} (F(\nu, (\xi_l))) = \lambda \pi  \frac{d\innpro{g_l, E_l (\nu) \xi_l}}{d \nu} = \lambda \pi  \frac{d\innpro{\xi_l, E_l (\nu) g_l}}{d \nu} \label{eq:general imaginary part}
\end{align}
for Lebesgue a.e. $\nu \in \bbR$.
The second term of (\ref{eq:Currents cal1}) has the following form:
\begin{align}
& {\rm Im} \innpro{F(h_{0,k}, (\xi_l)) g_k, \cbk{ (\calN_k(h_{0,k}) + 1/2) / \abs{\eta_-(h_{0,k})}^2 } g_k} \nonumber\\
=& - \lambda \pi \int_{\sigma_k} \frac{1}{\abs{\eta_-(\nu)}^2} \frac{d\innpro{g_l, E_l (\nu) \xi_l}}{d \nu} \rbk{\calN_k(\nu) + \frac{1}{2}} d\innpro{g_k, E_k(\nu) g_k} \label{eq:2nd term}
\end{align}

By equation (\ref{eq:def of linear form of PF weights}), we have that
\begin{align}
\innpro{ v_k, \varphi((1)) } = \frac{\lambda \innpro{v_k, g_k} }{\eta(0)}, \quad \innpro{v_k, \varphi_k((\xi_l))} = \delta_{k,l} \innpro{v_l, \xi_l} + \frac{\lambda^2 \innpro{v_k, g_k} \innpro{g_l, (h_{0,l})^{-1} \xi_l} }{\eta(0)}. \label{eq:3rd term}
\end{align}
Since $g_k \in \calD((h_{0,k})^{-1/2})$ for any $k = 1, \ldots, N$, $\eta(0)$ is finite and real valued.
By the equations (\ref{eq:1st term}), (\ref{eq:general imaginary part}), (\ref{eq:2nd term}), and (\ref{eq:3rd term}), we obtain
\begin{align}
& \omega_+(\Psi((1)) \Psi((i \xi_l))) - \omega_+(\Psi((i)) \Psi( (\xi_l))) \nonumber \\
=& 2 \pi \lambda^3 \sum_{k=1}^N \int_{\sigma_l} \frac{1}{ \abs{\eta_-(\nu)}^2 } \rbk{ \calN_l(\nu) - \calN_k(\nu) } \frac{d \innpro{g_k, E_k(\nu) g_k} }{d\nu} d\innpro{\xi_l, E_l(\nu) g_l} \nonumber\\
&  + \pi^3 \left[ \cbk{ \sum_{k = 1}^N \Theta_k \rbk{  \frac{\lambda \innpro{v_k, g_k} }{\eta(0)} } } \cbk{ \sum_{k=1}^N \Theta_k \rbk{ \delta_{k,l} i \innpro{v_l, \xi_l} + \frac{ \lambda^2 i \innpro{v_k, g_k} \innpro{g_l, (h_{0,l})^{-1} \xi_l} }{\eta(0)} } } \right. \nonumber\\
& \left. - \cbk{ \sum_{k=1}^N \Theta_k \rbk{  \frac{\lambda i \innpro{v_k, g_k} }{\eta(0)} }  }\cbk{ \sum_{k=1}^N \Theta_k\rbk{ \delta_{k,l} \innpro{v_l, \xi_l} + \frac{ \lambda^2 \innpro{v_k, g_k} \innpro{g_l, (h_{0,l})^{-1} \xi_l} }{\eta(0)} } } \right] \nonumber\\
=& 2 \pi \lambda^3 \sum_{k=1}^N \int_{\sigma_l} \frac{1}{ \abs{\eta_-(\nu)}^2 } \rbk{ \calN_l(\nu) - \calN_k(\nu) } \frac{d \innpro{g_k, E_k(\nu) g_k} }{d\nu} d\innpro{\xi_l, E_l(\nu) g_l} \nonumber\\
&  + \frac{\pi^3 \lambda}{\eta(0)} \sum_{k = 1}^N \cbk{ \Theta_k \rbk{ \innpro{v_k, g_k} } \Theta_l \rbk{ i \innpro{v_l, \xi_l} } - \Theta_k \rbk{   i \innpro{v_k, g_k} } \Theta_l \rbk{ \innpro{v_l, \xi_l} } } \nonumber\\
& + \frac{\pi^3 \lambda^3}{\eta(0)^2} \sum_{j, k=1}^N \cbk{ \Theta_j \rbk{ \innpro{v_j, g_j} } \Theta_k\rbk{ i \innpro{v_k, g_k} } - \Theta_j \rbk{  i \innpro{v_j, g_j}  } \Theta_k\rbk{ \innpro{v_k, g_k} } }\innpro{g_l, (h_{0,l})^{-1} \xi_l},
\end{align}
since $\xi_l = g_l$ or $h_{0,l} g_l$ and $\innpro{g_l, (h_{0,l})^{-1} \xi_l}$ is real valued.
If $\xi_l = g_l$ (resp. $\xi_l = h_{0,l} g_l$), then we have the equation (\ref{eq:general charge current}) (resp. (\ref{eq:general heat current})).\QED

For each $l=1, \ldots, N$, we define Josephson currents at NESS by
\begin{align}
{\rm Jos}_l(\omega_+) = \frac{\pi^3 \lambda^2}{\eta(0)} \sum_{k = 1}^N \cbk{ \Theta_k \rbk{ \innpro{v_k, g_k} } \Theta_l \rbk{ i \innpro{v_l, g_l} } - \Theta_k \rbk{   i \innpro{v_k, g_k} } \Theta_l \rbk{ \innpro{v_l, g_l} } }. \label{eq:Josephson current}
\end{align}
By using Corollary \ref{cor:general Currents}, we get an explicit formula of the mean entropy production rate.
\begin{pro} \label{pro:general Entropy production rate}
The mean entropy production rate $Ep(\omega_+)$ has the form of
\begin{align}
Ep(\omega_+) = \sum_{k,l=1}^N \int_{\sigma_l} \frac{\lambda^4 \pi }{\abs{\eta_-(\nu)}^2} \cbk{\beta_l (\nu - \mu_l) - \beta_k (\nu - \mu_k)} (\calN_k(\nu) - \calN_l(\nu)) \frac{d\innpro{g_k, E_k (\nu) g_k}}{d\nu}d \innpro{g_l, E_l (\nu) g_l}.
\end{align}
\end{pro}

\noindent
{\bf Proof.}
By Corollary \ref{cor:general Currents}, we have that
\begin{align}
& - \sum_{l=1}^N \beta_l \rbk{ \omega_+( E_l ) - \mu_l \omega_+( J_l ) } \nonumber\\
=& 2 \lambda^4 \pi \sum_{l=1}^N \sum_{k=1}^N \int_{\sigma_l} \frac{1}{\abs{\eta_-(\nu)}^2} (\beta_l \nu - \beta_l \mu_l) (\calN_k(\nu) - \calN_l(\nu))) \frac{ d\innpro{g_k, E_k (\nu) g_k} }{d \nu} d \innpro{g_l, E_l (\nu) \xi_l} \nonumber\\
& +  \frac{\pi^3 \lambda^2}{\eta(0)} \sum_{l=1}^N \beta_l \mu_l \sum_{k = 1}^N \cbk{ \Theta_k \rbk{ \innpro{v_k, g_k} } \Theta_l \rbk{ i \innpro{v_l, g_l} } - \Theta_k \rbk{   i \innpro{v_k, g_k} } \Theta_l \rbk{ \innpro{v_l, g_l} } }. \label{eq:gen ent prod rate}
\end{align}
If $\mu_l \neq 0$, then $\Theta_l \equiv 0$ and if $\Theta_l \not \equiv 0$, then $\mu_l=0$.
Thus, the last term of (\ref{eq:gen ent prod rate}) is equal to zero.
It follows that
\begin{align}
& - \sum_{l=1}^N \rbk{ \beta_l \omega_+( E_l ) - \beta_l \mu_l \omega_+( J_l ) } \nonumber\\
=& \sum_{l=1}^N \sum_{k = 1}^N \int_{\sigma_l} \frac{\lambda^4 \pi }{\abs{\eta_-(\nu)}^2} (\beta_l \nu - \beta_l \mu_l - \beta_k \nu + \beta_k \mu_k) (\calN_k(\nu) - \calN_l(\nu))) \frac{d\innpro{g_k, E_k (\nu) g_k}}{d\nu}d \innpro{g_l, E_l (\nu) \xi_l}. 
\end{align}
Thus, we have done.\QED

By the above proposition, the mean entropy production rate is independent of phase terms.
Thus, Josephson currents ${\rm Jos}_l(\omega_+)$ may occur without entropy production, if the temperatures and the chemical potentials of reservoirs are identical.

For any $k,l \in \{1, \ldots, N\}$, the function
\begin{align}
\frac{\lambda^4 \pi }{\abs{\eta_-(\nu)}^2} \frac{d\innpro{g_k, E_k (\nu) g_k}}{d\nu} \frac{d\innpro{g_l, E_l (\nu) g_l}}{d\nu}
\end{align}
corresponds to the ``total transmission probability'' (\cite{AschbacherJaksic}, \cite{Yafaev92}, and \cite{Yafaev10}).
As in \cite{AschbacherJaksic}, we say that the channel $k \to l$ is open if the set 
\begin{align}
\Set{ \nu \in \sigma_k \cap \sigma_l | \frac{1}{\abs{\eta_-(\nu)}^2}  \frac{d\innpro{g_k, E_k (\nu) g_k}}{d\nu}  \frac{d\innpro{g_l, E_l (\nu) g_l}}{d\nu} > 0 } \label{eq:Set}
\end{align}
has a positive Lebesgue measure.
If $\beta_k \neq \beta_l$ or $\mu_k \neq \mu_l$, then the function
\begin{align}
\cbk{\beta_l (\nu - \mu_l) - \beta_k (\nu - \mu_k)} (\calN_k(\nu) - \calN_l(\nu))
\end{align}
is strictly positive for any finite interval. 
Thus, we obtain the following theorem.

\begin{theorem} \label{theorem:positivity of EPR}
If there exists an open channel $k \to l$ such that either $\beta_k \neq \beta_l$ or $\mu_k \neq \mu_l$ for some $k,l \in \{ 1, \ldots, N\}$, then $Ep(\omega_+) > 0$.
\end{theorem}

\setcounter{theorem}{0}
\setcounter{equation}{0}
\section{Examples} \label{sec:examples}
In this section, we give calculations of currents on $\bbR^d$ and graphs.

\subsection{$L^2(\bbR^d)$, $d \geq 3$, case}
In this subsection, we put $\fraK_k = L^2(\bbR^d)$, $d \geq 3$, for any $k=1,\ldots, N$.
The case that a model consists of a quantum particle and two reservoirs is calculated in \cite{AccardiTasaki} and \cite{MatsuiTasaki}. 
We consider the case of the number of reservoirs more than two and the case of phase terms different from \cite{MatsuiTasaki}.
The Hamiltonians $h_{0,k}$ are Fourier transform of positive Laplacian on $L^2(\bbR^d)$, i.e. the multiplication operator of $\abs{p}^2/2$, $p \in \bbR^d$.
If $g_k \in C_0^\infty(\bbR^d)$ and $g_k$ are continuous with respect to $\abs{x}$ for any $k=1,\ldots,N$, then $g_k$ satisfies condition {\rm (A)}  and we have that
\begin{align}
\lim_{\varepsilon \searrow 0} {\rm Im} \innpro{g_k, (\nu - h_{0,k} - i\varepsilon)^{-1} g_k} = \left\{ \begin{array}{cc}
C(d) \nu^{\frac{d - 1}{2}} \abs{g(\sqrt{2\nu})}^2 & (\nu \geq 0) \\
0 & (\nu<0)
\end{array} \right. ,
\end{align}
where $C(d)$ is the dimension dependent constant.
Thus, we can find $\lambda > 0$ and $\Omega > 0$ such that the function $\eta(z)$, defined in (\ref{eq:definition of eta(z)}), satisfies condition {\rm (B)}.
We fix such $g$, $\lambda$, and $\Omega$.

Since the form factors $g_k$ are in $C^\infty_0(\bbR^d)$, there are compact sets $K_k \subset \bbR^d$ such that ${\rm supp} g_k \subset K_k$ for any $k = 1, \ldots, N$.
Let $\calK_1$ and $\calK_2$ be the Hilbert spaces defined by
\begin{align}
\calK_1 =  \bbC \oplus \rbk{ \bigoplus_{k=1}^N L^2(K_k) } , \quad \calK_2 = 0 \oplus\rbk{ \bigoplus_{k=1}^N L^2(\bbR^d \backslash K_k) }.
\end{align}
Note that $\calK = \calK_1 \oplus \calK_2$, $h \restriction_{\calK_2} = h_0 \restriction_{\calK_2}$, $h \calK_1 \subset \calK_1$, and $h \calK_2 \subset \calK_2$.
As a consequence, we have that
\begin{align}
e^{ith} = e^{ith} \restriction_{\calK_1} \oplus e^{ith_0} \restriction_{\calK_2}
\end{align}
on $\calK_1 \oplus \calK_2$.
Since $h \restriction_{\calK_1}$ and $h_{0} \restriction_{\calK_1}$ are bounded, we can use Lemma \ref{lmm:forms of terms of line_integrals} and obtain an explicit formula for $e^{ith}\restriction_{\calK_1}$.

For any $\psi, \xi \in C_0^\infty(\bbR^d)$, we have that $(\psi, \xi)$, $(\psi, g)$, and $(\xi, g)$ satisfy condition {\rm (Abs)} by using Mourre estimate techniques. (See e.g. \cite{Yafaev92} and \cite{Yafaev10}.)
Since $d \geq 3$, $C_0^\infty(\bbR^d) \subset \calD((h_{0,k})^{-1/2})$.
The PF weight for $\abs{x}^2/2$ is the delta function $\delta(x)$.
We put $\frah = \bigoplus_{k=1}^N C_0^\infty(\bbR^d)$.
For $k=1,\ldots, N$, we set $\mu_k = 0$ and 
\begin{align}
\Theta_k^{(1)}(\alpha) =& e^{i\tau_k} D_k^{1/2} \alpha + e^{-i\tau_k} D_k^{1/2} \overline{\alpha}, \label{eq:SSB}\\
\Theta_k^{(2)}(\alpha) =& s_{1,k} D_ k^{1/2} {\rm Re} \alpha + s_{2,k} D_k^{1/2} {\rm Im} \alpha, \label{eq:GCS}
\end{align}
where $\alpha \in \bbC$, $\tau_k \in [0, 2\pi)$, $D_k > 0$, $s_1, s_2 \in \bbR$ and $\overline{\alpha}$ is the complex conjugate for $\alpha$.
The terms $\Theta^{(1)}_k$ and $\Theta_k^{(2)}$ appear in a component of a factor decomposition of quasi-free states with BEC.
(See \cite[Section 5.2.5]{BratteliRobinsonII}, \cite[(1.18)]{Matsui2006}, and \cite[Theorem 4.5]{Kanda}.)
For $\psi_k \in C_0^\infty(\bbR^d)$, we define the initial states $\omega_0^{(1)}$ and $\omega_0^{(2)}$ by
\begin{align}
\omega_0^{(1)}(W((\psi_k))) = \exp \cbk{ -\frac{1}{4} \innpro{ \psi_k, (e^{\beta_k h_{0,k}} - \mathbbm{1})^{-1} \psi_k } + i \Theta^{(1)}_k \rbk{ \psi_k(0) } }, \\
\omega_0^{(2)}(W((\psi_k))) = \exp \cbk{ -\frac{1}{4} \innpro{ \psi_k, (e^{\beta_k h_{0,k}} - \mathbbm{1})^{-1} \psi_k } + i \Theta^{(2)}_k \rbk{ \psi_k(0) } }.
\end{align}
We assume that $\omega_0^{(1)}$ and $\omega_0^{(2)}$ satisfy condition {\rm (C)}.
Since the vectors $g, \psi \in \frah$ satisfy conditions {\rm (Abs)}, {\rm (A)}, {\rm (B)}, and {\rm (D)}, there are $\omega_+^{(1)}$ and $\omega_+^{(2)}$ and we have that
\begin{align}
\omega^{(1)}_+(J_l) =& 2 \pi \lambda^4 \sum_{k=1}^N \int_{\sigma_l} \frac{1}{ \abs{\eta_-(\nu)}^2 } \rbk{ \calN_l(\nu) - \calN_k(\nu) } \frac{d \innpro{g_k, E_k(\nu) g_k} }{d\nu} d\innpro{g_l, E_l(\nu) g_l} \nonumber\\
&  + \frac{4 \pi^3 \lambda^2}{\eta(0)} D_l^{1/2} \sum_{k = 1}^N D_k^{1/2} {\rm Im} \rbk{ e^{i (\tau_k - \tau_l)} g_k(0) \overline{g_l(0)} }, \label{eq:Jos current L^2 1} \\
\omega^{(2)}_+(J_l) =& 2 \pi \lambda^4 \sum_{k=1}^N \int_{\sigma_l} \frac{1}{ \abs{\eta_-(\nu)}^2 } \rbk{ \calN_l(\nu) - \calN_k(\nu) } \frac{d \innpro{g_k, E_k(\nu) g_k} }{d\nu} d\innpro{g_l, E_l(\nu) g_l} \nonumber\\
&  + \frac{\pi^3 \lambda^2}{\eta(0)} D_l^{1/2} \sum_{k = 1}^N D_k^{1/2} \cbk{ \rbk{s_{1, k} s_{1,l} + s_{2, k} s_{2, l} } {\rm Im} \rbk{ g_k(0) \overline{g_l(0)} } + \rbk{ s_{1, k} s_{2, l} - s_{1, l} s_{2, k} } {\rm Re}\rbk{ g_k(0) \overline{g_l(0)} } }. \label{eq:Jos current L^2 2}
\end{align}

\subsection{Graphs}
In this subsection, we give calculations of currents on periodic graphs and comb graphs.
To verify our conditions for the adjacency operators of undirected graphs, we apply results of M. M\u{a}ntoiu et al. \cite{MantoiuRichardAldecoa}.

Let $G = (VG, EG)$ be an undirected graph, where $VG$ is the set of vertices in $G$ and $EG$ is the set of edges in $G$.
Two vertices $x, y \in VG$ are said to be adjacent, if there exists an edge $(x,y) \in EG$ joining $x$ and $y$, and we write $x \sim y$.
For any $x \in VG$, we denote the degree of vertex $x \in VG$ by ${\rm deg}_G(x)$, i.e.
\begin{align}
{\rm deg}_G (x) = \sharp \set{ y \in VG | (x, y) \in EG },
\end{align}
and 
\begin{align}
{\rm deg}_G := \sup_{x \in VG} {\rm deg}_G(x).
\end{align}
In this paper, we assume that $G$ is connected with no loops and with no multiple edges, $VG$ is a countable set, and ${\rm deg}_G < \infty$. 
Let $\ell^2(VG)$ be the set of square summable sequence labeled by the vertices in $VG$.
Let $A_G$ be the adjacency operator of $G$ defined by
\begin{align}
\innpro{ \delta_x, A_G \delta_y } = \left\{ 
\begin{array}{cc}
1 & \text{ if } (x, y) \in EG \\
0 & \text{ if } (x,y) \not \in EG 
\end{array}
\right. .
\end{align}

\subsubsection{Adapted Graphs}
We recall the definition of adapted graphs introduced in \cite{MantoiuRichardAldecoa}.
For any $x \in VG$, we denote the set of neighbors of $x$ by $N_G(x)$, i.e. $N_G(x) := \set{ y \in VG | (x,y) \in EG }$.
\begin{df} {\rm (\cite[Definition 3.1.]{MantoiuRichardAldecoa})} \label{df:adapted}
A function $\Phi : G \to \bbR$ is adapted to the graph $G$ if the following conditions hold: 
\begin{description}
\item[{\rm (i)}] There exists $c \geq 0$ such that $\abs{ \Phi(x) - \Phi(y) } \leq c$ for any $x, y \in VG$ with $(x,y) \in EG$.
\item[{\rm (ii)}] For any $x, y \in VG$, one has
\begin{align}
\sum_{z \in N(x) \cap N(y)} \cbk{ 2 \Phi(z) - \Phi(x) - \Phi(y) } = 0.
\end{align}
\item[{\rm (iii)}] For any $x,y \in VG$, one has
\begin{align}
\sum_{z \in N(x) \cap N(y)} \cbk{ \Phi(z) - \Phi(x) } \cbk{ \Phi(z) - \Phi(y) } \cbk{ 2 \Phi(z) - \Phi(x) - \Phi(y) } = 0.
\end{align} 
\end{description}
A pair $(G, \Phi)$ is said to be adapted if $\Phi$ is adapted to $G$.
\end{df}

Let $(G, \Phi)$ be an adapted graph.
We define the unbounded multiplication operator $\Phi$ on $\ell^2(VG)$ by $(\Phi f)(x) = \Phi(x) f(x)$, $x \in VG$, where $f \in \ell^2(VG)$ with $\sum_{x \in VG} \Phi(x)^2 \abs{f(x)}^2 < \infty$.
We denote the domain of $\Phi$ by $\calD(\Phi)$.
We define the operator $K$ on $\ell^2(VG)$ by
\begin{align}
(K \xi)(x) := i \sum_{y \in N(x)} \cbk{ \Phi(y) - \Phi(x) } \xi(y), \quad \xi \in \ell^2(VG), \quad x \in VG. \label{eq:definition of K}
\end{align}
The operator $K$ is self-adjoint and bounded by condition (i) in Definition \ref{df:adapted}.
Note that $K$ and $A_G$ are commute.
Since $K$ is self-adjoint, we have that the orthogonal decomposition of $\ell^2(VG)$ by 
\begin{align}
\ell^2(VG) = \ker K \oplus \overline{{\rm ran} K},
\end{align}
where ${\rm ran} K$ is the range of $K$. 
We denote the restriction of $A_G$ onto $\overline{{\rm ran} K}$ by $A_{G, 0}$.
\begin{theorem} {\rm \cite[Theorem 3.3]{MantoiuRichardAldecoa}} \label{theorem:adapted absolutely continuity}
Let $(G, \Phi)$ be an adapted graph.
\begin{description}
\item[{\rm (i)}] For any $\xi \in {\rm ran} K \cap \calD(\Phi)$, there exists a constant $c_\xi > 0$ such that
\begin{align}
\sup_{\mu \in \bbR, \varepsilon > 0} \abs{ \innpro{\xi, (\mu - A_{G,0} \pm i \varepsilon)^{-1} \xi } } \leq c_\xi. \label{eq:boundedness condition}
\end{align}
\item[{\rm (ii)}] The operator $A_{G,0}$ has purely absolutely continuous spectrum.
\end{description}
\end{theorem}

\subsubsection{Radon--Nikodym Derivative for the Spectral Measure of the Adjacency Operators}
In this subsection, we collect facts of the Radon--Nikodym derivative of the adjacency operator of an adapted graph $(G, \Phi)$.
We use the same notation in the above subsection.
By Theorem \ref{theorem:adapted absolutely continuity}, we have the following lemma:
\begin{lmm} \label{lmm:bddness of form}
Let $(G, \Phi)$ be an adapted graph.
Then for any $\xi, \zeta \in {\rm ran} K \cap \calD(\Phi)$, there exists $c_{\xi, \zeta} > 0$ such that
\begin{align}
\sup_{\mu \in \bbR, \varepsilon > 0} \abs{ \innpro{\xi, (\mu - A_{G, 0} \pm i\varepsilon)^{-1} \zeta} } < c_{\xi, \zeta}.
\end{align}
\end{lmm}

\noindent
{\bf Proof.}
By polarization identity, we have that
\begin{align}
4 \innpro{\xi, (\mu - A_{G, 0} \pm i\varepsilon)^{-1} \zeta} = \sum_{k = 0}^3 (-i)^k \innpro{ (\xi + i^k \zeta), (\mu - A_{G, 0} \pm i\varepsilon)^{-1} (\xi + i^k \zeta)  }.
\end{align}
Since $\xi + i^k \zeta \in {\rm ran} K \cap \calD(\Phi)$, we obtain the statement by Theorem \ref{theorem:adapted absolutely continuity}.
\QED

By the above lemma, for any $\zeta, \xi \in {\rm ran} K \cap \calD(\Phi)$, the function $\innpro{ \xi, (z - A_{G, 0})^{-1} \zeta }$ is in $\bbH^\infty(\bbC_+)$.
\begin{lmm} \label{lmm:boundedness of angular limit}
For any $\xi, \zeta \in {\rm ran} K \cap \calD(\Phi)$, the function $f_{\xi, \zeta}(z) := \innpro{ \xi, (z - A_{G, 0})^{-1} \zeta }$ is in $\bbH^\infty(\bbC_+)$.
Moreover, for Lebesgue a.e $x \in \bbR$, the limit
\begin{align}
f^+_{\xi, \zeta}(x) := \lim_{\varepsilon \searrow 0} f_{\xi, \zeta}(x + i \varepsilon)
\end{align} 
exists and $f^+_{\xi, \zeta} \in L^\infty(\bbR, dx)$.
\end{lmm}

\noindent
{\bf Proof.}
Since $A_{G,0}$ is self-adjoint, $f_{\xi, \zeta}$ is holomorphic in $\bbC_+$. By \cite[Theorem 3.13]{Jaksic06} and Lemma \ref{lmm:bddness of form}, we have the statement.
\QED

For any $\xi, \zeta \in {\rm ran} K \cap \calD(\Phi)$, there exists the Radon--Nikodym derivative $d \innpro{ \xi, E(\nu) \zeta } / d\nu$, where $E$ is the spectral measure of $A_{G, 0}$.

\begin{lmm} \label{lmm:RN derivative property}
For any $\xi, \zeta \in {\rm ran} K \cap D(\Phi)$, Radon--Nikodym derivative $d \innpro{ \xi, E(\nu) \zeta } / d\nu$ exists and is in $L^p(\bbR)$ for any $p \in \bbN \cup \{ \infty \}$.
Moreover, $d \innpro{ \xi, E(\nu) \zeta } / d\nu$ is given by 
\begin{align}
\frac{d \innpro{ \xi, E(\nu) \zeta } }{ d\nu } = \lim_{ \varepsilon \searrow 0} \frac{1}{\pi} \cbk{ \innpro{ \xi, (\nu - A_{G,0} - i \varepsilon)^{-1} \zeta } - \innpro{ \xi, (\nu -A_{G,0} + i \varepsilon)^{-1} \zeta }}
\end{align}
for Lebesgue a.e. $\nu \in \bbR$.
The support of the function $d \innpro{ \xi, E(\nu) \zeta } / d\nu$ is contained in the set of the spectrum of $A_{G,0}$.
\end{lmm}

\noindent
{\bf Proof.}
Since the measure $d \innpro{\xi, E(\lambda) \zeta}$ is a finite complex measure and absolutely continuous with respect to Lebesgue measure,
by Radon--Nikodym theorem (see e.g. \cite[6. 10]{Rudin}), the Radon--Nikodym derivative $d \innpro{ \xi, E(\nu) \zeta } / d\nu$ is in $L^1(\bbR)$.
Note that for Lebesgue a.e. $\nu \in \bbR$,
\begin{align}
\frac{ d \innpro{ \xi, E(\nu) \zeta } }{ d\nu }= \frac{1}{\pi} \lim_{ \varepsilon \searrow 0} \cbk{ \innpro{ \xi, (\nu - A_{G,0} - i \varepsilon)^{-1} \zeta } - \innpro{ \xi, (\nu - A_{G,0} + i \varepsilon)^{-1} \zeta }}
\end{align}
by \cite[Theorem 4.15]{Jaksic06} and polarization identity.
Since $\innpro{ \xi, (\nu - A_{G,0} \pm i 0)^{-1} \zeta }$ is in $L^\infty(\bbR)$ by Lemma \ref{lmm:boundedness of angular limit}, $d \innpro{ \xi, E(\nu) \zeta } / d\nu$ is in $L^p(\bbR)$ for any $p \in \bbN \cup \{ \infty \}$.
If $\nu \not \in [ - \norm{A_G}, \norm{A_G} ]$, i.e. $\nu$ is in the resolvent set of $A_{G,0}$, then $(\nu - A_{G,0})^{-1}$ is bounded, and 
\begin{align}
\lim_{\varepsilon \to 0} (\nu - A_{G,0} \pm i \varepsilon)^{-1} = (\nu - A_{G,0})^{-1}
\end{align}
in norm sense.
Thus, we have the statement.
\QED

\subsubsection{$\bbZ^d$, $d \geq 3$, case}
In this subsection, we consider $\bbZ^d$, $d \geq 3$, as graphs.
We note that $\bbZ^d$ has an adapted function $\Phi$ defined by
\begin{align}
\Phi((x_1, \ldots, x_d)) = \sum_{k=1}^d x_k, \label{eq:example of adapted function}
\end{align}
where $x_k \in \bbZ$.
Then the operator $K$ defined in (\ref{eq:definition of K}) has the form of
\begin{align}
K \delta_{x} = i \sum_{k=1}^d (\delta_{x - e_k} - \delta_{x + e_k}),
\end{align}
where $x \in \bbZ^d$ and $e_k$ is the element of $\bbZ^d$ and the $k$-th component of $e_k$ is equal to $1$ and otherwise equal to $0$.
We put $\fraK_k = \ell^2(\bbZ^d)$, $h_{0,k} = \norm{A_{\bbZ^d}} \mathbbm{1} - A_{\bbZ^d}$, and 
\begin{align}
\frah_k := {\rm span} \set{ e^{ith_{0,k}} \delta_x | t \in \bbR, x \in \bbZ^d } 
\end{align}
for any $k = 1, \ldots, N$.
For example, we set
\begin{align}
g_k = K \delta_{x_k} = i  \sum_{j=1}^d (\delta_{x_k - e_j} - \delta_{x_k +  e_j} )
\end{align}
for some $x_k \in \bbZ^d$.
By Theorem \ref{theorem:adapted absolutely continuity}, $g_k$ satisfies condition (A).
Using the Fourier transformation, we have that
\begin{align}
\lim_{\varepsilon \searrow 0} {\rm Im} \innpro{ g_k, (\nu - h_{0,k} - i\varepsilon)^{-1} g_k } = (2 \pi)^{-d/2} \pi \int_{\bbT^d} \delta(\nu - \sum_{j=1}^d \sin^2 (\theta_j / 2)) \abs{\sum_{j = 1}^d \sin \theta_j}^2 d\theta. \label{eq:Spectral measure on Z^d}
\end{align} 
We put $\lambda^2 > 0$ such that $\lambda^2 C_g << 1$ and there exists $\Omega \in (0, \norm{h_{0,0}})$ such that the right hand side of (\ref{eq:Spectral measure on Z^d}) has some strictly positive lower bound for any $\nu \in [\Omega - \lambda^2 C_g, \Omega + \lambda^2 C_g]$.
Thus, condition {\rm (B)} are satisfied.
Since $h_{0,k}$ are transient, i.e. 
\begin{align}
\sup_{x \in \bbZ^d} \innpro{ \delta_x, (h_{0,k})^{-1} \delta_x } < \infty,
\end{align}
by $d \geq 3$, the form factor $g$ satisfies condition {\rm (D)}.
Note that $\bigoplus_{k=1}^N (\frah_k \cap {\rm ran} K) \subset \frah(g)$ by Lemma \ref{lmm:bddness of form}, where $\frah (g)$ is the set defined in (\ref{eq:def of abs subset}).
For an initial state $\omega_0$ with condition {\rm (C)} and with the form of (\ref{eq:initial state}), there exists NESS $\omega_+$ which is a state on $\calW(\frakk, \sigma)$, where $\frakk = \bbC \oplus \rbk{ \bigoplus_{k=1}^N \rbk{ \frah_k \cap {\rm ran}K } }$.
If the PF weight $v$ is defined by $v(x) = 1$ for any $x \in \bbZ^d$, then $\innpro{ v_k, g_k } = 0$ for any $k = 1, \ldots, N$.
Thus, ${\rm Jos}_l(\omega_+) = 0$ for any $l=1, \ldots, N$.

\subsubsection{Regular Admissible Graphs}
A graph $G$ is called regular, if for any $x, y \in VG$, ${\rm deg}_G(x) = {\rm deg}_G(y)$.
Recall the definition of admissible graphs (cf. \cite{MantoiuRichardAldecoa}).
In this subsection, we assume that $G$ is deduced from a directed graph, i.e., some relation $<$ is given on $G$ such that, for any $x, y \in VG$, $x \sim y$ is equivalent to $x < y$ or $y < x$, and one can not both $y < x$ and $x < y$.
We also write $y > x$ for $x < y$.
Then for any $x \in VG$, the neighbor of $x$, $N_G(x)$,  decompose into a disjoint union $N_G(x) = N^+_G(x) \cup N^-_G(x)$, where 
\begin{align}
N_G^+(x) := \set{ y \in VG | x < y }, \quad N_G^-(x) := \set{ y \in VG | y < x }. 
\end{align}
When directions have been fixed, we use the notation $(G, <)$ for the directed graph and say that $(G, <)$ is subjacent to $G$.

Let $p = x_0 x_1 \cdots x_n$ be a path.
We define the index of path $p$ by
\begin{align}
{\rm ind}(p) := \abs{ \set{ j | x_{j-1} < x_j } } - \abs{ \set{ j | x_{j-1} > x_j} }.
\end{align}

\begin{df} {\rm \cite[Definition 5.1]{MantoiuRichardAldecoa}}
A directed graph $(G,<)$ is called admissible if
\begin{description}
\item[{\rm (i)}] It is univoque, i.e., any closed path in $G$ has index zero.
\item[{\rm (ii)}] It is uniform, i.e., for any $x, y \in X$, $\sharp \rbk{ N^-_G(x) \cap N^-_G(y) } = \sharp \rbk{ N^+_G(x) \cap N^+_G(y) }$.
\end{description}
A graph $G$ is called admissible if there exists an admissible directed graph $(G, <)$ subjacent to $G$.
\end{df}

\begin{df} {\rm \cite[Definition 5.2]{MantoiuRichardAldecoa}}
A position function on a directed graph $(G, <)$ is a function $\Phi : G \to \bbZ$ satisfying $\Phi(x) + 1 = \Phi(y)$ if $x < y$.
\end{df}

\begin{lmm} {\rm \cite[Lemma 5.3]{MantoiuRichardAldecoa}}
\begin{description}
\item[{\rm (i)}] A directed graph $(G, <)$ is univoque if and only if it admits a position function.
\item[{\rm (ii)}] Any position function on an admissible graph $G$ is surjective.
\item[{\rm (iii)}] A position function on a directed graph $G$ is unique up to constant.
\end{description}
\end{lmm}

\begin{rmk}
If $G$ is an admissible graph, then there exists a position function $\Phi$.
The function $\Phi$ satisfies Definition \ref{df:adapted}.
Thus, an admissible graph is an adapted graph as well.
\end{rmk}

\begin{rmk}
When $G$ is an infinite regular graph, we only consider the PF weight $v$ for the adjacency operator $A_G$ defined by $v(x) = 1$ for any $x \in VG$.
\end{rmk}

\begin{pro} \label{pro:zero PF}
Let $G$ be an admissible regular graph.
Assume that $g \in \frah \cap {\rm ran} K$, where $\frah$ is defined by
\begin{align}
\frah := {\rm span} \set{ e^{it (\norm{A_G} \mathbbm{1} - A_G)} \delta_x | x \in VG, t \in \bbR }. \label{eq:def of subspace}
\end{align} 
Then $\innpro{v, g} = 0$.
\end{pro}

\noindent
{\bf Proof.}
Note that $\frah \subset \calD(v)$ by \cite[Theorem 4.5]{Fidaleo}.
Since $g \in \frah \cap {\rm ran} K$, there exists $\zeta \in \ell^2(VG)$ such that $g = K \zeta$, where $K$ is the operator defined in (\ref{eq:definition of K}).
Then the vector $g$ has the form of
\begin{align}
\innpro{\delta_x, g} = i \sum_{y \in N^+_G(x)} \zeta(y) - i \sum_{y \in N^-_G(x)} \zeta(y),
\end{align}
where $\zeta(y) = \innpro{\delta_y, \zeta}$.
Then we have that
\begin{align}
\innpro{v, g} = i \cbk{ \sum_{x \in VG} \sum_{y \in N^+_G(x)} \zeta(y) - \sum_{x \in VG} \sum_{y \in N_G^-(x)} \zeta(y) } = 0.
\end{align}
Thus, we have done.\QED

By the above proposition, we have the following theorem:
\begin{theorem} \label{theorem:admissible regular}
Let $G_k$, $k=1, \ldots, N$, be admissible regular graphs.
Fix $g \in \bigoplus_{k=1}^N \rbk{ \frah_k \cap {\rm ran} K_k }$, where $\frah_k$ are defined in {\rm (\ref{eq:def of subspace})} and $K_k$ are the operators defined in {\rm (\ref{eq:definition of K})}.
For any $k=1, \ldots, N$, we assume that $h_{0,k} = \norm{A_{G_k}}\mathbbm{1} - A_{G_k}$ is transient, there exist $\Omega, \lambda > 0$ such that the function $\eta(z)$ defined in {\rm (\ref{eq:definition of eta(z)})} satisfies condition {\rm (B)}, and the initial state $\omega_0$ satisfies condition {\rm (C)}.
Then there exists NESS $\omega_+$ which is a state on $\calW(\frakk, \sigma)$, where $\frakk = \bbC \oplus \bigoplus_{k=1}^N \rbk{ \frah_k \cap {\rm ran} K_k }$.
Moreover, for any $l=1,\ldots,N$, we have that
\begin{align}
\omega_+(J_l) =& 2 \pi \lambda^4 \sum_{k=1}^N \int_{\sigma_l} \frac{1}{ \abs{\eta_-(\nu)}^2 } \rbk{ \calN_l(\nu) - \calN_k(\nu) } \frac{d \innpro{g_k, E_k(\nu) g_k} }{d\nu} d\innpro{g_l, E_l(\nu) g_l},
\end{align}
where $J_l$ is defined in {\rm (\ref{eq:obs of charge})}.
\end{theorem}

\noindent
{\bf Proof.}
By assumptions, Theorem \ref{theorem:general Non-equilibrium steady states}, Corollary \ref{cor:general Currents}, and Proposition \ref{pro:zero PF}, we have the statement.\QED

\subsection{Comb Graphs}
In this subsection, we consider typical example of non-regular graphs: comb graphs.
BEC on comb graphs studied in \cite{Burioni}, \cite{FidaleoGuidoIsola}, and \cite{Fidaleo}.
In \cite{Burioni}, R. Burioni et al. calculated the spectral measure of the adjacency operators of comb graphs $\bbZ^d \dashv \bbZ$.
Using their results, we calculate currents on comb graphs.
First, we recall the definition of comb graphs.
\begin{df}
Let $G_1$ and $G_2$ be graphs, and let $o \in VG_2$ be a given vertex.
Then the comb product $X := G_1 \dashv (G_2, o)$ is a graph with vertex $VX := VG_1 \times VG_2$, and $(g_1, g_2), (g^\prime_1, g_2^\prime) \in VX$ are adjacent if and only if $g_1 = g_1^\prime$ and $g_2 \sim g_2^\prime$ or $g_2 = g_2^\prime = o$ and $g_1 \sim g_1^\prime$. We  call $G_1$ the base graph, and $G_2$ the fiber graph.
When $o \in VG_2$ is understood from the context, we omit it, and write $G_1 \dashv G_2$.
\end{df}
We consider the graphs $G_d := \bbZ^d \dashv (\bbZ, 0)$, $d \geq 3$.
As the case of $\bbZ^d$, the function $\Phi$ defined in (\ref{eq:example of adapted function}) is adapted to $G_{d-1}$.
Put $h_{0,l} = \norm{A_{G_d}} \mathbbm{1} - A_{G_d}$ for any $l = 1, \ldots, N$.
For $J \in \bbZ^d$ and $x \in \bbZ$, the operators $K_l$, $l = 1, \ldots, N$, have the form of 
\begin{align}
K \delta_{J, x} = \left\{ \begin{array}{cc}
i \delta_{J, x - 1} - i \delta_{J, x + 1} & (x \neq 0) \\
i \sum_{l = 1}^d \rbk{ \delta_{J - e_l, x} - \delta_{J + e_l, x} } + i \delta_{J, -1} - i \delta_{J, 1} & (x = 0)
\end{array} \right. .
\end{align}
Put $g_l = K \delta_{J_l, x_l}$, $l = 1, \ldots, N$, where $J_l \in \bbZ^d$ and $x_l \in \bbZ$ with $\abs{x_l} >> 1$.
Then, by Theorem \ref{theorem:adapted absolutely continuity} and \cite[Theorem 10.14]{FidaleoGuidoIsola}, the form factor $g$ satisfies conditions {\rm (A)} and {\rm (D)}.
By \cite[Lemma 9.4]{FidaleoGuidoIsola}, a PF weight $v$ has the following form:
\begin{align}
v(J, x) = \frac{e^{- \abs{x} \theta_d}}{2 \norm{ (2 \sqrt{d^2 + 1} - A_{\bbZ})^{-1} \delta_0 } \sinh \theta_d}, \quad J \in \bbZ^d, \quad x \in \bbZ,
\end{align}
with $2 \cosh \theta_d = 2 \sqrt{d^2 + 1}$.
Another example of $v$ is given in \cite{Fidaleo}.
The form of the spectral measure of $A_{G_d}$ is in \cite{Burioni}.
Thus, we can find $\Omega, \lambda > 0$ which satisfy condition {\rm (B)}.

The pairing of $g_l$ and the PF weight $v_l = v$ has the form of
\begin{align}
\innpro{v, g_l} =& i \frac{e^{- \abs{x_l - 1} \theta_d}}{2 \norm{ (2 \sqrt{d^2 + 1} - A_{\bbZ})^{-1} \delta_0 } \sinh \theta_d} - i \frac{e^{- \abs{x_l + 1} \theta_d}}{2 \norm{ (2 \sqrt{d^2 + 1} - A_{\bbZ})^{-1} \delta_0 } \sinh \theta_d} \nonumber\\
=& i  \frac{e^{- \abs{x_l} \theta_d}}{\norm{ (2 \sqrt{d^2 + 1} - A_{\bbZ})^{-1} \delta_0 }}.
\end{align}
Thus, we define the initial states $\omega^{(1)}_0$ and $\omega_0^{(2)}$ with condition {\rm (C)} by
\begin{align}
\omega_0^{(1)}(W((\psi_k))) = \exp \cbk{ -\frac{1}{4} \innpro{ \psi_k, (e^{\beta_k h_{0,k}} - \mathbbm{1})^{-1} \psi_k } + i \Theta^{(1)}_k \rbk{ \innpro{v_k, \psi_k} } }, \label{eq:def of initial state 1}\\
\omega_0^{(2)}(W((\psi_k))) = \exp \cbk{ -\frac{1}{4} \innpro{ \psi_k, (e^{\beta_k h_{0,k}} - \mathbbm{1})^{-1} \psi_k } + i \Theta^{(2)}_k \rbk{ \innpro{v_k, \psi_k} } } \label{eq:def of initial state 2}
\end{align}
for $\psi_k \in \frah_k \cap {\rm ran} K$, where $\Theta_k^{(1)}$ defined in (\ref{eq:SSB}) and $\Theta_k^{(2)}$ defined in (\ref{eq:GCS}).
Note that $\frakk := \bigoplus_{k=1}^N (\frah_k \cap {\rm ran} K) \subset \frah(g)$ by Lemma \ref{lmm:bddness of form}.
Thus, there exists NESS $\omega_+$, which is a state on $\calW(\frakk, \sigma)$.
If the temperatures are identical, then Josephson currents have the form of
\begin{align}
\omega_+^{(1)}(J_l) =& {\rm Jos}_l(\omega_+^{(1)}) = \frac{4 \pi^3 \lambda^2}{\eta(0)} \sum_{k = 1}^N D_k^{1/2} D_l^{1/2} \sin (\tau_k - \tau_l) \frac{e^{-(\abs{x_k} + \abs{x_l}) \theta_d}}{ \norm{(2 \sqrt{ d^2 + 1} - A_\bbZ)^{-1} \delta_0}^2 }, \\
\omega_+^{(2)}(J_l) =& {\rm Jos}_l(\omega_+^{(2)}) = \frac{\pi^3 \lambda^2}{\eta(0)} \sum_{k = 1}^N D_k^{1/2} D_l^{1/2} \cbk{ s_{1, k} s_{2, l} - s_{1,l} s_{2,k} } \frac{e^{-(\abs{x_k} + \abs{x_l}) \theta_d}}{ \norm{(2 \sqrt{ d^2 + 1} - A_\bbZ)^{-1} \delta_0}^2 }.
\end{align}

\section*{Acknowledgments}
The author would like to thank Professor Taku Matsui for discussions and comment of this work.


\normalsize

\end{document}